\documentclass[sigconf]{acmart}

\AtBeginDocument{%
  }

\copyrightyear{2025} 
\acmYear{2025} 
\setcopyright{cc}
\setcctype{by}
\acmConference[CHI '25]{CHI Conference on Human Factors in Computing Systems}{April 26-May 1, 2025}{Yokohama, Japan}
\acmBooktitle{CHI Conference on Human Factors in Computing Systems (CHI '25), April 26-May 1, 2025, Yokohama, Japan}
\acmDOI{10.1145/3706598.3713589}
\acmISBN{979-8-4007-1394-1/25/04}

\usepackage{acmart-taps}
\usepackage{xcolor}
\usepackage{tabularray}
\usepackage{subcaption}
\usepackage{enumitem}
\usepackage{booktabs}   
\usepackage{array}      
\usepackage{arydshln}   
\usepackage{caption}   
\usepackage{color}

\begin{document}

%%
%% The "title" command has an optional parameter,
%% allowing the author to define a "short title" to be used in page headers.
% \title{TutorUp: A Scenario-Based Learning Environment to Facilitate the Adoption of Effective Engagement Strategies in Remote Tutoring}
\title[TutorUp: Training Tutors to Address Engagement Challenges in Online Learning]{TutorUp: What If Your Students Were Simulated? Training Tutors to Address Engagement Challenges in Online Learning}
% %%
% %% The "author" command and its associated commands are used to define
% %% the authors and their affiliations.
% %% Of note is the shared affiliation of the first two authors, and the
% %% "authornote" and "authornotemark" commands
% %% used to denote shared contribution to the research.
% \author{Ben Trovato}
% \authornote{Both authors contributed equally to this research.}
% \email{trovato@corporation.com}
% \orcid{1234-5678-9012}
% \author{G.K.M. Tobin}
% \authornotemark[1]
% \email{webmaster@marysville-ohio.com}
% \affiliation{%
%   \institution{Institute for Clarity in Documentation}
%   \city{Dublin}
%   \state{Ohio}
%   \country{USA}
% }

\author{Sitong Pan}
\affiliation{%
  \institution{Zhejiang University}
  \city{Hangzhou}
  \country{China}}
\email{pstong@zju.edu.cn}

\author{Robin Schmucker}
 \affiliation{%
 \institution{Carnegie Mellon University}
 \city{Pittsburgh}
 \state{PA}
 \country{USA}
}
\email{rschmuck@cs.cmu.edu}

\author{Bernardo Garcia Bulle Bueno}
\affiliation{%
\institution{Massachusetts Institute of Technology}
\city{Cambridge}
\state{MA}
\country{USA}}
\email{bernard0@mit.edu}

\author{Salome Aguilar Llanes}
\affiliation{%
\institution{Massachusetts Institute of Technology}
\city{Cambridge}
\state{MA}
\country{USA}}
\email{saloagui@mit.edu}

\author{Fernanda Albo Alarcón}
\affiliation{%
  % \institution{Jóvenes Ayudando a Niñas y Niños}
  \institution{Autonomous Technological Institute of Mexico}
  \city{Mexico City}
  \country{Mexico}}
\email{fer.albo.al@gmail.com}

 \author{Hangxiao Zhu}
\affiliation{%
   \institution{Texas A\&M University}
   \city{College Station}
   \state{TX}
   \country{USA}
   }
\email{hangxiao@tamu.edu}

 \author{Adam Teo}
\affiliation{%
   \institution{Texas A\&M University}
   \city{College Station}
   \state{TX}
   \country{USA}
   }
\email{adamt321@tamu.edu}

\author{Meng Xia}
\affiliation{%
   \institution{Texas A\&M University}
   \city{College Station}
   \state{TX}
   \country{USA}
   }
\email{mengxia@tamu.edu}

% %%
% %% By default, the full list of authors will be used in the page
% %% headers. Often, this list is too long, and will overlap
% %% other information printed in the page headers. This command allows
% %% the author to define a more concise list
% %% of authors' names for this purpose.
\renewcommand{\shortauthors}{Pan et al.}

%%
%% The abstract is a short summary of the work to be presented in the
%% article.
\begin{abstract}
With the rise of online learning, many novice tutors lack experience engaging students remotely. We introduce \textit{TutorUp}, a Large Language Model (LLM)-based system that enables novice tutors to practice engagement strategies with simulated students through scenario-based training. Based on a formative study involving two surveys ($N_1=86$, $N_2=102$) on student engagement challenges, we summarize scenarios that mimic real teaching situations. To enhance immersion and realism, we employ a prompting strategy that simulates dynamic online learning dialogues. \textit{TutorUp} provides immediate and asynchronous feedback by referencing tutor-students online session dialogues and evidence-based teaching strategies from learning science literature. In a within-subject evaluation ($N=16$), participants rated \textit{TutorUp} significantly higher than a baseline system without simulation capabilities regarding effectiveness and usability. Our findings suggest that \textit{TutorUp} provides novice tutors with more effective training to learn and apply teaching strategies to address online student engagement challenges. 
\end{abstract}
% helps tutors learn and apply strategies for more effective student engagement, contributing to improved practices in online tutoring.
%%
%% The code below is generated by the tool at http://dl.acm.org/ccs.cfm.
%% Please copy and paste the code instead of the example below.
%%



%%
%% Keywords. The author(s) should pick words that accurately describe
%% the work being presented. Separate the keywords with commas.

%\received{20 February 2007}
%\received[revised]{12 March 2009}
%\received[accepted]{5 June 2009}

%%
%% This command processes the author and affiliation and title
%% information and builds the first part of the formatted document.
\maketitle

\section{Introduction}
Online tutoring and learning have become an increasingly important and widely adopted form of education~\cite{paudel2021online,doi:10.3102/0034654315581420,dumford2018online,10.1145/3613904.3641985}. Specifically, the outbreak of the COVID-19 pandemic was a pivotal moment accelerating the transition to online learning, allowing people to observe its inherent benefits, scalability, and flexibility for online education~\cite{aristovnik2023impact, thomas2023so, whenthetutorbecomes}. This shift has led to broader acceptance and adoption of online learning which demands a larger amount of online-teaching tutors~\cite{barno2024scaling,whenthetutorbecomes,thomas2023so,paudel2021online}. To fill this gap, more novice tutors and part-time tutors are recruited. Many of these new online-teaching tutors are not professionally trained~\cite{thomas2023so}, and may lack sufficient knowledge to handle certain situations with students, such as managing technical issues, fostering student engagement, and planning courses~\cite{vlachopoulos2021quality}.
  
This trend has prompted researchers to explore innovative approaches to train and prepare tutors for the challenges of online learning environments~\cite{thomas2023so,rosenberg2021addressing}. A prominent method is scenario-based tutor training, which simulates real-world teaching scenarios to provide educators with practical training opportunities~\cite{developmentofscenario}. Notable examples include clinical simulations proposed by \citet{dotger2013had} and the PLUS system developed by \citet{personalizedlearning}. The emergence of large language models (LLMs) has introduced more effective and intelligent methods for scenario-based tutor training~\cite{jin2024teach,gpteach, lee23generative}. Systems like GPTeach~\cite{gpteach} and studies by \citet{lee23generative} have demonstrated the potential of LLMs in simulating realistic training scenarios for tutor development. While these works highlight the feasibility and benefits of scenario-based training, they fall short in addressing the critical aspect of designing and authoring targeted scenarios, such as those focusing on student engagement challenges. Therefore, we aimed to utilize LLMs to develop an effective scenario-based training system tailored for novice tutors to practice engaging students in online learning.

To identify the pressing needs of online tutors and inform the design of effective scenario-based training solutions, we conducted a formative study comprising of two surveys. The first survey included $N_1 = 86$ tutors, and the second survey included $N_2 = 102$ tutors, both from JANN\footnote{https://jann.mx}. JANN is an existing online learning platform that connects thousands of volunteer tutors with K-12 students for free online math sessions in Mexico. The first survey aimed to investigate the challenges tutors encounter when teaching online. The results revealed that \textbf{engagement issues} stand out to be the most significant challenge with tutors expressing a strong need for strategies and training to address this problem effectively.
To gain a deeper insight into novice tutors' problems with student engagement, we performed the second survey with open-ended questions which aimed to understand the specific scenarios in which student disengagement manifests and how tutors have been addressing these challenges. We used a thematic analysis method~\cite{joffe2011thematic} to analyze the result and identified four themes to represent different forms of student disengagement: \textit{Lack of Interest and Engagement}, \textit{Lack of Confidence}, \textit{Varying Learning Speeds}, and \textit{Fatigue and Focus Issues}. The findings from our two rounds of surveys provided valuable guidance for designing a scenario-based training system specifically targeting student engagement issues.

Based on these survey results, the literature review on the effectiveness of scenario-based methods~\cite{clark2009accelerating,preservice,grossman2008back,adapting,scenariobased,prospectsforchange} and the feasibility of using large language models (LLMs) to simulate students~\cite{gpteach}, we designed our system, \textit{TutorUp}, as illustrated in Fig. ~\ref{fig:system}. Focusing on providing training for novice tutors to address student engagement problems, \textit{TutorUp} leverages GPT-4o~\cite{openai2023chatgpt4} to simulate student conversations, presenting common scenarios of engagement challenges in online learning. Tutors can interact with simulated students by typing instructions, which provides a realistic teaching scenario for practice. Additionally, we provide both immediate and asynchronous feedback to help novice tutors practice engaging students with effective strategies. 

To assess the usefulness and usability of \textit{TutorUp}, we conducted a within-subjects user study with $16$ participants, who are novice tutors conducting remote tutoring.  
The contributions of this work are summarized as follows:
\begin{itemize}
\item \textbf{Survey of Challenges in Remote Tutoring}: We present findings from a formative study involving two rounds of surveys ($N1 = 86$, $N2 = 102$) with tutors teaching online. We identify student engagement as a central factor for learning success, outline common scenarios where disengagement can occur, and curate teaching strategies based on tutor feedback and review of learning science literature.
\item \textbf{LLM-based System for Practicing Engagement Strategies}: Based on the requirements identified from our survey, we design and evaluate \textit{TutorUp}—a training system that aids tutors in learning strategies to promote student engagement. The system offers scenario-based training by simulating real teaching situations using LLM-based student agents and provides feedback referencing user inputs and established teaching strategies.
\item \textbf{User Study with Online Tutors}: We present results from a within-subject user study ($N = 16$) comparing \textit{TutorUp} to a baseline system. Participants rated \textit{TutorUp} significantly higher in terms of effectiveness and usability. Evaluations of conversational transcripts suggest that \textit{TutorUp} improved the acquisition and application of engagement strategies.
\end{itemize}

\section{Related Work}
\label{sec:related_worl}

\subsection{Scenario-Based Tutor Training}

Scenario-based training is an educational paradigm advocating for practice in authentic contexts that simulate real-world situations~\cite{preservice}.
For tutor education, scenario-based learning offers a solution to a common challenge faced by novice teachers--\textit{learning passively from observing others with limited opportunities for active practice of teaching strategies in low-stakes settings}~\cite{prospectsforchange}.
%  before starting their practicum
Scenarios that are authentic and aligned with real-world contexts enable learners to build strong connections with applications they will encounter in their professional lives~\cite{teachingearly}.
% Simulations facilitate effective scenario creation.
A prominent example is clinical simulations, where participants respond to scenarios they are likely to face in their profession~\cite{dotger2013had,teachermoments}. This method has been utilized in healthcare for decades and has since been widely adopted in the field of tutor education~\cite{teachermoments,simulationbasedlearning}. \citet{dotger2013had} and~\citet{designingandusing} both developed simulations in which trained actors assume the roles of ``representative'' characters (e.g., a parent or student). These actors engage with participating educators and reliably introduce specific, scripted dialogue and themes into the simulation~\cite{dotger2013had,clinicalsimulations}. However, despite the value of such simulations, they are challenging to organize and costly to implement. To address these issues,~\citet{teachermoments} developed Teacher Moments as a digital version of simulations. The platform offers a cost-effective and efficient way to simulate scenarios, enabling tutors to repeatedly practice various teaching situations. The PLUS platform, developed by Lin et al.~\cite{developmentofscenario, whenthetutorbecomes,usinglarge,personalizedlearning}, recently advanced scenario-based simulated courses to train tutors in addressing math difficulties and motivational barriers. However, their scenarios are written as text, and tutors only provide solutions but cannot interact with students in the designed scenarios.
These previous studies have demonstrated the effectiveness of scenario-based training. However, more work is needed to investigate what typical engagement issue scenarios are and how to utilize LLMs to simulate the scenarios and provide tutors with feedback to address engagement issues in online learning. This paper develops a tutor training system where scenarios are simulated by intelligent systems (LLMs) and are designed with theories and insights from learning science literature~\cite{casestudy, abou2021emergency, enhancing,sevenprinciples,motivation} and a formative study with dozens of ($N_1=86$, $N_2=102$) real tutors. 

\subsection{LLM-based Human Simulation and Feedback Generation in Education}
LLMs like ChatGPT-4o~\cite{openai2023chatgpt4} have proven highly effective in simulating human interactions and activities in educational contexts, including classroom scenarios~\cite{zhang2024simulating, thomas2023so, ma2024students}, learning by teaching~\cite{jin2024teach, schmucker2023ruffle}, and feedback generation~\cite{hirunyasiri2023comparative, lin2024can, dai2024assessing, stamper2024enhancing, demszky2024does}. 
 
Our work develops a scenario-based tutor training system and leverages LLMs to 1) simulate the interactions between students and tutor and 2) provide personalized feedback for tutors to practice.

In terms of using LLMs to simulate students' interactions, a significant work is SimClass~\cite{zhang2024simulating}, a multi-agent classroom simulation. SimClass deploys multiple agents with distinct roles, such as teacher, assistant, and classmates, to simulate a dynamic classroom environment, enhancing collaborative learning experiences among virtual students. \citet{lee23generative} integrated LLMs into immersive problem-solving environments, showing their effectiveness in simulating realistic student personas and challenges. Similarly, GPTeach~\cite{gpteach} is a chat-based tool allowing novice tutors to interact with GPT-based simulated students exhibiting varied personas, learning goals, and engagement levels. However, these works including GPTeach lack feedback for more effective training practice and improvement.

LLMs have been widely applied to assess tutors' open-text responses and provide personalized feedback both immediately~\cite{aleven2006toward} and asynchronously~\cite{dai2024assessing, stamper2024enhancing}. This offers an efficient alternative to the traditionally time-intensive process of feedback generation~\cite{dai2024assessing,stamper2024enhancing}. Researchers have employed various  techniques, including few-shot prompting~\cite{fewshot} and chain-of-thought (CoT) prompting~\cite{chain} to provide qualified feedback. For example,~\citet{hirunyasiri2023comparative} explored zero-shot and few-shot CoT prompting to provide timely feedback on tutors' use of praise for students~\cite{hirunyasiri2023comparative}. Notably,~\citet{learningandAI} employed zero-shot learning and prompt chaining to provide feedback specifically in scenario-based tutor training~\cite{learningandAI}, showing that LLMs can deliver high-quality feedback on tutors' scenario-based responses. These studies collectively highlight the effectiveness of LLMs in providing feedback for tutor training. 
% Our work builds upon these foundations to further advance the application of LLMs in this context.

% Studies on using LLMs for simulating human interactions and generating feedback demonstrate the feasibility and effectiveness of creating an LLM-based tutor training system. 
Building on these foundations, we integrate LLMs into our system, \textit{TutorUp}, for scenario simulation and feedback generation, creating a realistic and accessible training system that enhances tutors' ability to engage students in online learning environments.

\subsection{Measuring Engagement in Online Learning Environments}
\label{subsec:engagement}

To provide feedback on engagement issues, we surveyed works on engagement definition and measurement. Engagement is recognized as a complex and multifaceted concept, with its precise definition varying across contexts and domains~\cite{Hagerup1993, joshi2022behavioral, redmond2018online,fredricks2004school}. A prevalent definition by~\citet{fredricks2004school} breaks down engagement into three distinct dimensions, capturing emotional, behavioral, and cognitive aspects. In the context of online learning, scholars added dimensions of collaborative engagement, focusing on peer interaction, as well as social engagement, facilitated through teaching strategies promoting community and sense of belonging~\cite{joshi2022behavioral,redmond2018online}. 

Numerous approaches have been designed to measure learners' engagement levels, based on the diverse ways in which learners display behaviors, convey emotions, and direct cognitive efforts towards learning tasks.~\citet{booth2023engagement} have comprehensively reviewed engagement measurement methods, categorizing them into traditional and modern methods. Traditional methods rely on human-driven assessments, including retrospective self-reports (e.g., questionnaires, interviews)~\cite{turner2000studying,gorin2001recall}, momentary self-reports (e.g., experience sampling~\cite{csikszentmihalyi1987validity, hutt2019time}), and observer-based measures (e.g., video coding~\cite{zaletelj2017predicting,yun2018automatic,sumer2021multimodal}, or live observations like the BROMP protocol~\cite{ocumpaugh2015baker}). While effective, traditional methods face challenges such as scalability, time intensity, and potential human biases. In contrast, modern methods utilize machine learning and sensor technologies to dynamically infer engagement from features such as video~\cite{bidwell2011classroom,dhall2018emotiw}, audio~\cite{dhall2018emotiw}, and interaction logs~\cite{grawemeyer2017affective,dewan2019engagement}. This automation enhances scalability, reduces costs, and enables real-time interventions, addressing limitations of traditional methods~\cite{booth2023engagement}.

In this work, with the guidance of previous learning theories on engagement~\cite{fredricks2004school, joshi2022behavioral}, we explore how to use LLMs to automatically analyze simulated students' engagement changes by examining their conversational responses and provide feedback from four dimensions: emotional, behavioral, cognitive, and collaborative.
\section{Formative Study: Understanding Challenges and Scenarios of Engagement in Remote Tutoring}
\label{sec:formative_study}

\subsection{Study Context}

We conducted a formative study including two rounds of surveys to gain a deeper understanding of the specific challenges tutors face in online teaching (first survey) and inform the design of effective scenario-based training
solutions (second survey). From the initial survey, we identified managing student engagement as the biggest challenge. Focusing on this issue, the second survey investigated the scenarios of student disengagement and how tutors address them. All survey questions are provided in Appendix \ref{apdx:sureveyquestions}. 

We worked closely with tutors on the JANN platform to complete surveys for this formative study. JANN is a non-profit online platform that matches volunteers with groups of kids to teach them. Tutors receive no payment, and the majority are college students who volunteer to fulfill a social service requirement for obtaining a university degree in Mexico. The requirements to become a tutor on this platform are minimal: a brief interview (1-3 minutes) and completion of a basic teaching training (3 hours). As a result, tutors' experience varies widely, with many having no prior teaching experience. This makes effective training tools and materials essential for JANN, which strongly motivates our work.

\aptLtoX{\begin{table*}
\centering
\caption{Reactive Scenario Themes. Based on tutors' descriptions of how engagement related challenges surface in their classes, we simulate scenarios allowing tutors to practice teaching strategies promoting student engagement.}
\label{tab:reactiv_scenarios} % Label for referencing the table
\begin{tabular}{p{150pt}p{150pt}p{150pt}}
\toprule
Scenario Theme                  & Description                                                                                              & Reactive Scenario                                                         \\
\midrule
Lack of Interest and Engagement & Students are disengaged in class, showing little interest or participation in learning activities.       & Students don't attend the class~                                          \\
                                &                                                                                                          & Students don't respond to asked questions/group messages~                 \\
                                &                                                                                                          & Students don't show reaction to the taught content (never ask questions)~ \\
                                &                                                                                                          & Students not paying attention to the lecture content                      \\
                                &                                                                                                          & Students did not want to participate                                      \\
                                \hline
Lack of Confidence              & Students lack confidence in their abilities, which affects their participation and performance in class. & Students have low self-condidence                                      \\
                                &                                                                                                          & Students are afraid to make mistakes                                                   \\
                                &                                                                                                          & Students know the answer to something but don't reply~                    \\
                                \hline
Varying Learning Paces          & Students have varying learning speeds, with faster learners feeling bored and slower ones struggling to keep up.                & Some students learn faster and get bored                                  \\
                                &                      & Students making very basic errors like subtraction or multiplication      \\
                                \hline
Fatigue and Focus Issues        & Students appear tired and unable to concentrate, affecting their learning and engagement.                & Students appear tired in the class                                        \\
\bottomrule
\end{tabular}
\end{table*}}{\begin{table*}
\centering
\caption{Reactive Scenario Themes. Based on tutors' descriptions of how engagement related challenges surface in their classes, we simulate scenarios allowing tutors to practice teaching strategies promoting student engagement.}
\label{tab:reactiv_scenarios} % Label for referencing the table

\begin{tblr}{
  width = \linewidth,
  colspec = {Q[35]Q[45]Q[60]},
  row{1} = {c},
  cell{2}{1} = {r=5}{},
  cell{2}{2} = {r=5}{},
  cell{7}{1} = {r=3}{},
  cell{7}{2} = {r=3}{},
  cell{10}{1} = {r=2}{},
  cell{10}{2} = {r=2}{},
  hline{1-2,7,10,12-13} = {-}{},
}
Scenario Theme                  & Description                                                                                              & Reactive Scenario                                                         \\
Lack of Interest and Engagement & Students are disengaged in class, showing little interest or participation in learning activities.       & Students don't attend the class~                                          \\
                                &                                                                                                          & Students don't respond to asked questions/group messages~                 \\
                                &                                                                                                          & Students don't show reaction to the taught content (never ask questions)~ \\
                                &                                                                                                          & Students not paying attention to the lecture content                      \\
                                &                                                                                                          & Students did not want to participate                                      \\
Lack of Confidence              & Students lack confidence in their abilities, which affects their participation and performance in class. & Students have low self-condidence                                      \\
                                &                                                                                                          & Students are afraid to make mistakes                                                   \\
                                &                                                                                                          & Students know the answer to something but don't reply~                    \\
Varying Learning Paces          & Students have varying learning speeds, with faster learners feeling bored and slower ones struggling to keep up.                & Some students learn faster and get bored                                  \\
                                &                      & Students making very basic errors like subtraction or multiplication      \\
Fatigue and Focus Issues        & Students appear tired and unable to concentrate, affecting their learning and engagement.                & Students appear tired in the class                                        
\end{tblr}
\end{table*}}

\subsection{Survey 1: Overall Challenges in Online Tutoring}
We conducted a web-based survey investigating the challenges JANN's tutors face.
The survey included two open-ended questions, focusing on the challenges they faced in the past and present.

We received responses from 86 tutors, including 47 females and 39 males, of whom 17 are full-time college students. Tutors' average age was 24 years and median age was 22 years.
%They provided valuable insights into the challenges of online teaching.
Two researchers conducted a thematic analysis~\cite{joffe2011thematic} of the survey responses using the Affinity Diagramming~\cite{kawakita1991original} method to classify and interpret the data. Specifically, one researcher performed the initial classification of all responses, while the other reviewed the categorization to ensure accuracy. When encountering conflicts, the reviewer labeled them and the two researchers discussed their interpretations and referred to the original responses until they reached a consensus. After the classification, they also discussed the names of the categories. 

From the challenges identified in the responses, we highlighted issues such as class preparation, tutor confidence, and maintaining engagement. Among these we found that maintaining student engagement is the most crucial issue, closely tied to factors such as scheduling (e.g., students not attending), tutor skills and mindset (e.g., nervousness and lack of practice), and the teaching process (e.g., keeping students' attention). Overall, these analysis results motivate us to design a tutor training system with strategies to help tutors address these student engagement challenges.

\subsection{Survey 2: Scenarios of Engagement Issues in Online Learning}
\label{subsec:survey2}

Through a literature review, we identified scenario-based training as an effective method to train tutors as it could provide tutors with realistic scenario simulation for practicing~\cite{teachingearly, prospectsforchange}. Thus, we conducted the second survey specifically focusing on identifying the exact scenarios tutors faced and the strategies they employed to manage student engagement. We designed three open-ended questions asking about tutors' observations of student engagement and their strategies for addressing it. To make our survey rigorous and effective, we conducted a comprehensive review of the relevant literature on student engagement~\cite{abou2021emergency,motivation,enhancing,sevenprinciples,casestudy}. From this review, we identified and summarized potential strategies for mitigating engagement problems, which we then categorized into ten distinct categories (e.g., show empathy to students and set academic goals) with specific instances. The full summary of the strategy can be found in Appendix \ref{apdx:strategies}. Based on these strategies, we formulated specific questions for tutors, inquiring whether they have tried these strategies and, if so, how and when they implemented them.

The questions in this second round of the survey are thus structured into two main parts:
\begin{enumerate}
    \item \textbf{Open-ended Questions:}
    \begin{enumerate}%[label=\textbf{Q\arabic*:}]
        \item[\textbf{Q1:}] \textit{What types of student engagement problems have you encountered?}
        \item[\textbf{Q2:}] \textit{Why do you think students are disengaged?}
        \item[\textbf{Q3:}] \textit{What strategies have you used to increase students' engagement and when did you use them?}
    \end{enumerate}
    \item \textbf{Strategy-specific Questions:}
    \begin{enumerate}%[label=\textbf{Q\arabic*:}, resume]
        \item[\textbf{Q4:}] \textit{Have you employed a specific strategy (e.g., set time limitation)?}
        \item[\textbf{Q5:}] \textit{If yes, how and when did you implement this strategy?}
    \end{enumerate}
\end{enumerate}

We received valid responses from 102 tutors on JANN, including 60 females and 42 males. The average age was 27 years and the median age was 24 years. Additionally, 91 had no prior online tutoring experience before they started at JANN, while the other 17 had an average of 18 hours of past tutoring experience. 

The survey analysis involved two main steps, identifying and matching scenario-strategy pairs and clustering scenarios using affinity diagramming ~\cite{kawakita1991original}, conducted by the same two researchers from the first survey. First, scenarios were identified from the responses and matched with corresponding strategies. One researcher summarized the scenarios and matched all responses, while the second reviewed them. For open-ended questions ($N=3$), scenarios involving student disengagement and related strategies were identified to form scenario-strategy pairs. For strategy-specific questions ($N=10$), scenarios were extracted based on the provided strategies to form scenario-strategy pairs. For example, a response to the strategy-specific question ``Did you discuss behavioral expectations with your student?'' such as ``The student did not want to participate'' was matched to form the pair: ``Student did not want to participate - Behavioral expectations.'' In cases of conflict, the researchers first discussed their differing interpretations and reviewed the original responses until consensus was reached. Finally, the scenario-strategy pairs were grouped and labeled to identify common scenarios and strategies, a classification step similar to the first survey. This analysis resulted in summarized scenario types with corresponding strategies to address each scenario. For example, in a scenario type where students appear tired in class, matched strategies such as ``explaining to reduce students' burden'' and ``adding attractive activities'' can be used to address the issue. % and thus are matched to it 

From the analysis result, we identified that some scenarios were specifically centered on student-related factors, such as knowledge level (e.g., fast learners getting bored), personality traits (e.g., unconfident, shy), or classroom behavior (e.g., tiredness, don't respond). Tutors addressed these situations by implementing strategies tailored to the specific disengagement issues of students.
Others were more general, focusing on broader classroom contexts, such as ``at the beginning'' or ``always'' where teachers proactively maintained engagement without targeting specific student behaviors. 
% \deleted{Based on the responses, we categorized both the scenarios and the corresponding strategies. We found that the responses primarily addressed two types of scenarios and their associated strategies: \textit{preventive scenarios} and \textit{reactive scenarios}.} 
These two types of scenarios align with the Proactive-Reactive behavioral patterns in student-teacher relationships~\cite{yucel2010analysis}. Based on this alignment, we formulated the scenarios and their associated strategies into two overarching types:
    
\textbf{Proactive Scenarios}: \textit{Tutors implement strategies to proactively maintain engagement across various contexts.} In the survey, when answering the ``how and when do you use this strategy'' questions, some tutors provided general scenarios, such as ``at the beginning'', ``whenever students are asked to do exercises,'' or even ``always.'' We observed that these strategies were not aimed at addressing specific disengagement issues but at preventing disengagement and promoting overall student engagement to maintain focus.

\textbf{Reactive Scenarios}: \textit{Tutors take strategies to address specific instances of disengagement as they occur.} In contrast to proactive strategies, some tutors reported taking specific measures in response to particular student issues. For example, in response to the question ``Why do you think students are disengaged?'', tutors mentioned scenarios such as "some students learn faster and get bored" or ``students know the answer but don't reply.'' In these cases, tutors reacted to students' behaviors and employed targeted strategies to promote engagement. From tutors' responses, we identified reactive situations where specific disengagement issues arose. Out of 1,326 responses (102 participants answering 13 questions), 138 were identified as describing reactive scenarios. 
In this analysis process, reactive scenarios were categorized into four themes: \textit{1) lack of interest and engagement, 2) lack of confidence, 3) varying learning paces, 4) fatigue and focus issues.} Tab.~\ref{tab:reactiv_scenarios} shows description and examples of reactive scenarios for each theme. 

Overall, reactive disengagement scenarios are more challenging for novice tutors, as they require instant and context-specific interventions~\cite{park2022frustration}. Therefore, reactive scenarios were the primary focus of this study.

\subsection{Design Requirements:}
\subsubsection{General Requirements}
Here we summarize key insights from the formative study, emphasizing the need for a tutor training system focused on student engagement in online learning. To ensure the system is effective and constructive, we propose two main design requirements to support tutors:
\begin{enumerate}%[label=R\arabic*]
    \item[R1] \textbf{Simulated Online Learning Scenarios}: \\
    The system should simulate realistic online learning scenarios, allowing tutors to practice their teaching skills in an environment where reactive situations are presented.

    \item[R2] \textbf{Feedback and Improvement Strategies}: \\
    The system should provide feedback to assess tutors' performances and offer insights and strategies for improvements.
\end{enumerate}

\subsubsection{Design Iteration}
\label{subsec:iteration}
We invited five experienced Math tutors from JANN (2 male, 3 female, aged 24 to 30, with an average age of 26.6) to participate in design iteration interviews to gather their insights and suggestions for updating our system. Their online tutoring experience ranged from 4 to 6 months, with 3 having additional teaching backgrounds in private tutoring and formal school environments (average 3.3 years).  These interviews were conducted via Zoom, where tutors used our system under our instruction. When trying our system, they were asked to think aloud and express their ideas. Follow-up questions on students and feedback design were also included. They generally found the simulated scenarios realistic and the feedback valuable for practicing engagement strategies. However, they suggested improvements including that the strategies matched to the scenarios were incomplete. We asked them to help refine the strategy matching using the ten categories identified in Section \ref{subsec:survey2}, which improved the scenario-strategy alignment (see the complete results in supplementary materials). We also integrated other valuable feedback to enhance system effectiveness and user experience (details in Section \ref{sec:system_design}).

\section{System Design and Implementation}
\label{sec:system_design}

\begin{figure*}
    \centering
    \includegraphics[width=\textwidth]{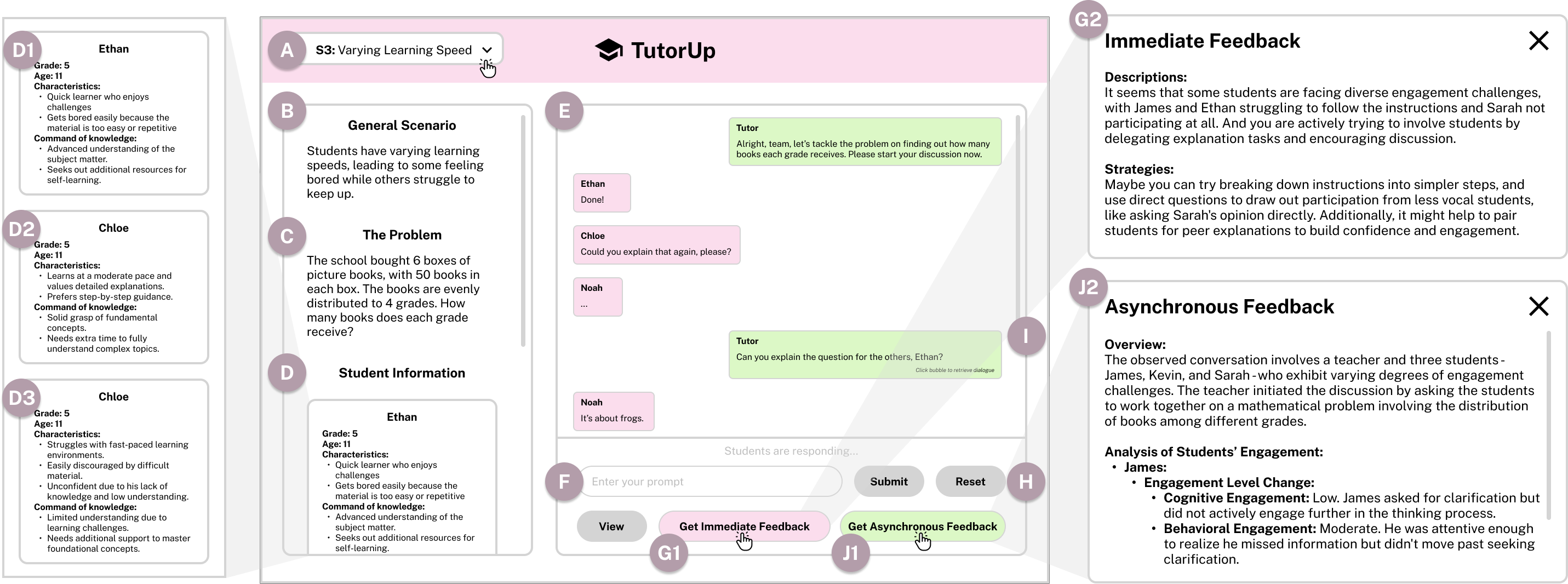}
    \caption{\textbf{TutorUp} consists of multiple views and functions, including Scenario Selection (A), Scenario Description (B), Math Problem (C), and Student Information (D), which displays details for three students (D1, D2, and D3) under each scenario. The system also features an Interactive Dialogue (E), an Input Box for Tutor Instructions (F), a Get Immediate Feedback Button (G1) that provides Immediate Feedback (G2), a Reset Button (H), and a Get Asynchronous Feedback Button (J1) that provides Asynchronous Feedback (J2). Users can also click on Tutor Dialogue Bubbles To Retrieve the Instruction (I)}
    \label{fig:system}
\end{figure*}

\subsection{System Usage Introduction}
We designed \textit{TutorUp}  (Fig.~\ref{fig:system}), a training system to help novice tutors address student engagement challenges in online tutoring. 
Users begin by selecting one of four scenarios (summarized in Subsection \ref{subsec:survey2} and listed in Tab. \ref{tab:reactiv_scenarios}) representing different disengagement issues (Fig. \ref{fig:system} A). Each scenario includes a description of the disengagement problem (Fig. \ref{fig:system} B) and a mathematical problem (Fig. \ref{fig:system} C) which serves as session context. Tutors aim to enhance student engagement through discussions based on the scenario information.

Before instructing the simulated students, tutors can review their information in the current scenario (Fig.~\ref{fig:system} D), including their grade level, age, personal characteristics (e.g., anxious about making mistakes), and their level of prior knowledge (e.g., limited understanding due to learning difficulties). These student profiles align with the general scenario description provided above (Fig. \ref{fig:system} A). The simulated students are age 10-11 years (4th-5th grade), an age group identified through iterative design (Subsection \ref{subsec:iteration}) as particularly prone to engagement issues.

After reviewing the scenario information, tutors gain an understanding of the students' situations and discussion topic. They initiate the online session by responding to an initial prompt in the dialogue window (Fig. ~\ref{fig:system} E). Tutors enter their instructions in the text box (Fig. \ref{fig:system} F), and the simulated students respond to the tutor and peers accordingly. Tutors' primary task is to engage the students, with the mathematical problem as context. The focus for tutors is on fostering student engagement rather than correct answers, as a student may give the correct answer but still be disengaged, for example, by guessing or repeating a peer’s answer without genuine involvement.

While interacting with the simulated students, whenever the tutor is in need—e.g., when they feel that their current strategy is ineffective—they can click on the \textit{Get Immediate Feedback} button (Fig. ~\ref{fig:system} G1). This provides immediate feedback (Fig. ~\ref{fig:system} G2) with an analysis of the classroom situation and suggested teaching strategies. Tutors can choose to implement the suggested strategies or continue with their teaching approach. To try different strategies, tutors can click the Reset button (Fig.~\ref{fig:system} H) to return to the initial dialogue or click on a previous dialogue bubble (Fig.~\ref{fig:system} I) to revert to an earlier state in the conversation.

Once tutors feel they have practiced enough, they can end the conversation and click the \textit{Get Asynchronous Feedback} button (Fig. ~\ref{fig:system} J1) to receive a comprehensive assessment of their teaching (Fig.~\ref{fig:system} J2). This feedback evaluates their performance and offers suggestions for improvement. Tutors can then apply these insights in further practice to test and refine their strategies.

To build the \textit{TutorUp} system, we developed three key components: first, we summarized typical disengagement scenarios from survey results and designed student personas to represent each scenario. Second, we implemented the BigPicture-Character prompting pipeline to simulate tutoring discussions with multiple students. These two components meet \textbf{R1}. Third, we utilized LLMs to review these dialogues to give immediate feedback and asynchronous evaluations for tutors to reference, which meets \textbf{R2}. The following subsections will introduce these three components in detail. The full prompts used for scenario simulation and feedback generation with LLMs, as described in the following subsections, can be found in the supplementary materials.

%%%%%%%%%%%%%%%%%%%%%%%%%%%%%%%%%%%%%%%%%%%%%%%%%%%%%%%%%%%%%%%%%%%%%%%%%%%%%%%%
\subsection{Scenario Simulation Design}
\label{subsec:senario_design}

Each simulated scenario features a scenario theme, student information, and a math problem. We discussed scenario themes in Subsection \ref{subsec:survey2} and show them in Tab. \ref{tab:reactiv_scenarios}. This section introduces the design for students under each scenario and the Math Problems.

\subsubsection{Student Design}
Based on the scenario themes (Tab. ~\ref{tab:reactiv_scenarios}), we designed student personas for each disengagement theme. Fig.\ref{fig:system} shows an example for the \textit{Varying Learning Paces} scenario. Details for all four scenarios are in the supplementary materials. From prior literature~\cite{gpteach,li2024agent} and survey responses (Section ~\ref{sec:formative_study}), we identified four key components for student personas: name, characteristics, command of knowledge, and initial behaviors. During the iterative design process (Subsection \ref{subsec:iteration}) tutors further suggested adding age and grade to help them adjust expectations regarding the tone and content of conversations, which we incorporated. The characteristics cover the student's personality (e.g., shyness), attitude (e.g., finding academic topics dull), emotions (e.g., fear of peer judgment), and tendencies (e.g., second-guessing themselves). Command of knowledge captures varying learning abilities (e.g., a good grasp of concepts) and preferences (e.g., a quick learner who prefers challenging problems). Initial behavior represents student's early disengagement signs (e.g., unwillingness to answer), which may change based on the tutor's instructions. This behavior is not displayed in the interface but is revealed through conversational interactions.
For each scenario, personas emphasize key traits related to the theme. For example, in the \textit{Lack of Interest and Engagement} scenario, the personas primarily highlight disinterested traits through their characteristics, while maintaining similar levels of knowledge.

\subsubsection{Math Problems}
To enhance realism and provide context, we designed algebraic math problems with simple structures and fixed answers for tutors and students to discuss. These problems allow tutors to focus on managing student engagement without additional complexity. In our iterative design interviews (Subsection \ref{subsec:iteration}), tutors confirmed that simple math problems are effective for general education and can be tied to real-world scenarios (e.g., counting apples), making them useful for guiding students.  Here is an example problem in our system: ``\textit{A farmer has two types of fruit trees: apple trees and pear trees. The total number of fruit trees on the farm is 120. The number of apple trees is twice the number of pear trees. Your task is to find out how many apple trees and pear trees there are on the farm.}'' We designed four similar problems, which can be combined with any scenario theme. Descriptions of all problems are in the supplementary materials.

\subsection{Generation of LLM-simulated Scenario}
To simulate scenarios for online tutoring sessions, \textit{TutorUp} formulates specific prompts to send API calls to GPT-4o. Throughout system development and evaluation we set the temperature parameter to $0$ to make system outputs more predictable. Future iterations will consider higher temperatures to introduce variations within scenarios supporting tutors in repeated practice. The prompt scheme for simulation in \textit{TutorUp} consists of two core components: one prompts LLMs to simulate individual student behavior, and another to govern interactions among students and the tutor, ensuring coherence and realism in the online learning environment.

\begin{figure*}
    \centering
    \includegraphics[width=\textwidth]{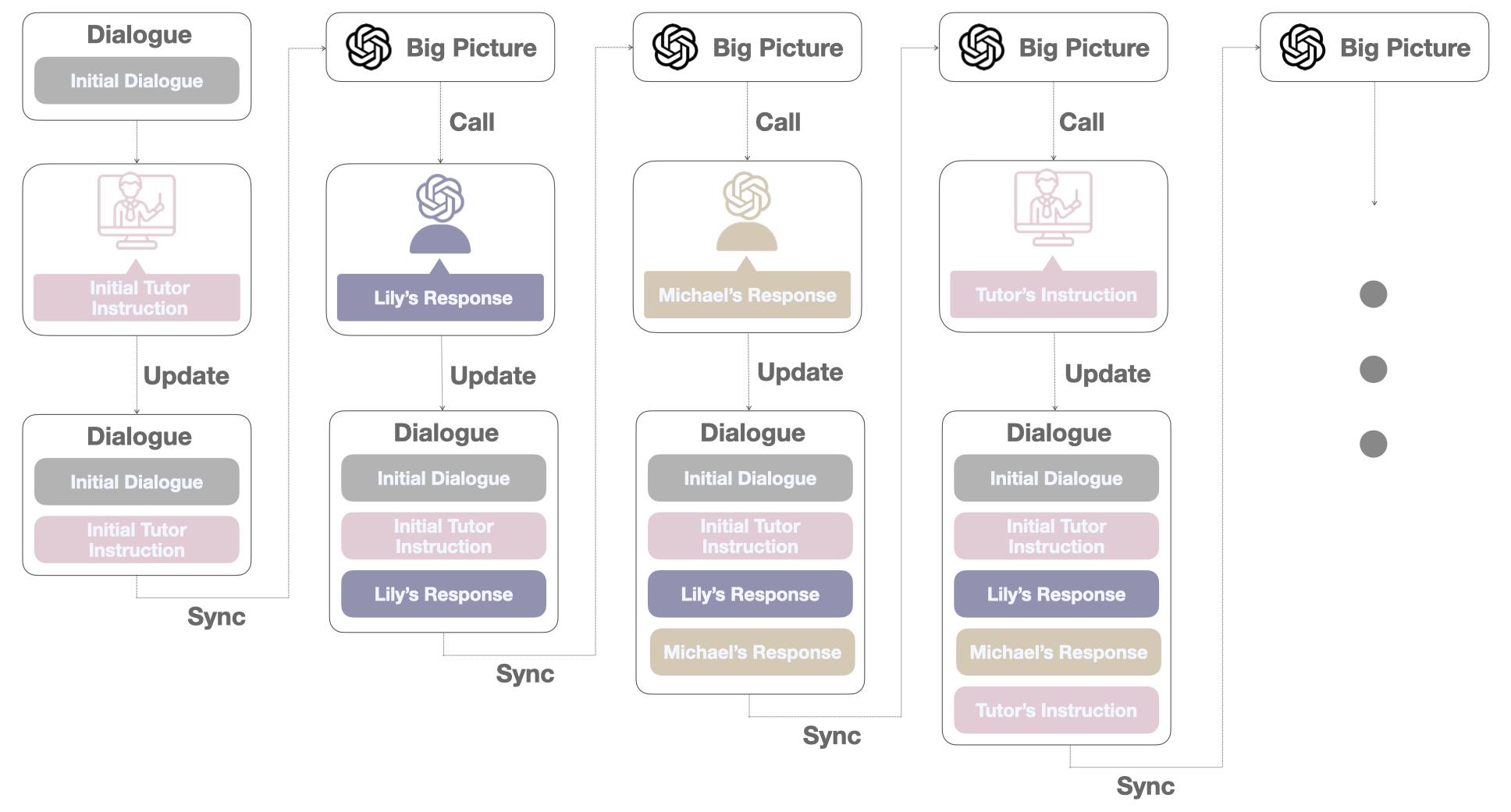}
%award level: inter school
\caption{BigPicture-Character Pipeline Demonstration: Beginning with the initial dialogue and tutor's first input, the BigPicture Agent starts to write the ``story'' of the dialogue and decides which Character (Tutor or Student Agent) to speak next.}
    \label{fig:promptPipeline}
\end{figure*}

\subsubsection{Generation of LLM-simulated Student Agent}
To accurately simulate an online discussion environment, the simulated students should closely align with real student behaviors in our specifications (Subsection ~\ref{subsec:senario_design}) and their dialogue should closely resemble that of real students in similar settings. To achieve this, we developed a \textit{student prompt} template, which models student personas and generate dialogues in the online context. The template was developed through multiple experiments. Initially, for example, we defined student behavior in the input prompt but this led to simulated students remaining disengaged, regardless of the tutor's instructions. Instead, we revised the prompt to establish an initial behavior and allow students' responses, which reflect their engagement, to evolve according to the tutor's guidance. Feedback from design interviews (Subsection \ref{subsec:iteration}) led to further updates, such as ensuring students maintain polite and respectful language. We kept the best-performing template. Below, we describe the main components of the prompt design:
\begin{enumerate}
    \item \textbf{Scenario Context Prompt}: \texttt{Please assume the role of a disengaged student in an online mathematics class. Your objective is to exhibit disengaged behavior, making it challenging for the tutor to engage you. The problem to be solved is: [INTRODUCE THE MATH PROBLEM].}
    \item \textbf{Student Information Prompt}: \texttt{Your name is [STUDENT NAME]. You are a [STUDENT AGE]-year-old student. Communicate in a manner appropriate for your age; you are not exceptionally quick to understand the problem. You possess characteristics of [STUDENT CHARACTERISTICS], and your understanding of the math problem is [STUDENT ABILITY]. Your initial disengaged behavior is [STUDENT BEHAVIOR]. Please respond politely to the tutor's greetings and questions. Your engagement level starts low but may increase if the tutor's strategies are effective.}
    \item \textbf{Output Requirements Prompt}: \texttt{Your responses should be brief, natural, and reflective of a disengaged student. You must maintain politeness and respect in all interactions with the tutor and classmates.}
\end{enumerate}

%%%%%%%%%%%%%%%%%%%%%%%%%%%%%%%%%%%%%%%%%%%%%%%%%%%%%%%%%%%%%%%%%%%%%%%%%%%%%%%%
\subsubsection{BigPicture-Character Prompting Pipeline}
While student agents simulate individual student dialogue, the online scenario requires both student-to-student and student-to-tutor interactions. We addressed four key questions: Q1: When should students speak? Q2: When should the tutor provide instructions? Q3: Which students should respond? Q4: In what order should student agents speak? We found that using a single agent to manage both the flow and content of the dialogue risks losing context and coherence over longer dialogues easily. To address this, we separated the responsibilities, with the BigPicture Agent focusing on managing the overall conversational flow and assigning speaking turns, while the Student Agents handle the individual student responses. We introduce the \textit{BigPicture-Character prompt} scheme below.

\textbf{BigPicture-Character Design Consideration:} It is assumed that each student can think independently and respond naturally to both tutor instructions and input from other students. A natural dialogue among students and tutor can be treated as a story. 
 
Inspired by previous work that utilizes LLMs to tell stories~\cite{mirowski2023co,shakeri2021saga,grigis2024playwriting}, we utilize this method to comprehensively manage the dialogue. We adopt a ``divide-and-conquer'' approach by introducing a BigPicture Agent to manage the overall flow of the conversation and determine the speaking order. If the speaker is a student, the corresponding Student Agent is called to respond; if it is a tutor, the system pauses and waits for the tutor's instruction. The BigPicture Agent focuses solely on determining which character will speak next. The specific content of each character's dialogue is decided by the student agent or the tutor's input. We include the scenario context, the profiles of three students, and the math problem as part of the big-picture agent's prompt input to ensure that the generated dialogue follows a logical order and interaction pattern~\cite{stamper2024enhancing}. 

\textbf{Dialogue Flow under the Picture-Character Prompt Pipeline:} As can be seen in Fig. \ref{fig:promptPipeline}, the BigPicture Agent maintains a dialog list that starts with an initial dialog, which is shared with each Student Agent. When the Tutor gives guidance for the first time, Tutor's speech will be added to both the BigPicture and Student Agent dialog lists, updating the scene’s ``story.'' Next, The system calls the BigPicture Agent to continue the ``story'' and determine the next character to speak. In this example, the BigPicture Agent assigns the next turn to Lily, a Student Agent, who generates a response based on the updated dialog. Lily's response is then added to the dialog lists of the BigPicture and all Student Agents. Afterward, the BigPicture Agent continues the conversation, pointing to James's speech, which is also updated in the log. The BigPicture Agent then continues with the Tutor's speech, and instead of calling other agents, the system waits for the Tutor's input, updating all dialog records accordingly. This pipeline creates a natural, interactive dialog between Tutor and Students. Refer to Fig.~\ref{fig:promptPipeline} for the detailed prompt pipeline corresponding to the example described above.

%%%%%%%%%%%%%%%%%%%%%%%%%%%%%%%%%%%%%%%%%%%%%%%%%%%%%%%%%%%%%%%%%%%%%%%%%%%%%%%%
\subsection{Feedback Design}
\textit{TutorUp} combines GPT-4o~\cite{openai2023chatgpt4} with evidence-based teaching strategies to provide personalized feedback for tutors interacting with simulated students. Two types of feedback are provided: Immediate and Asynchronous feedback~\cite{wong2017using}, each with different timing and content. Below are design considerations for both.

\subsubsection{Immediate Feedback} Immediate Feedback provides contextually relevant and personalized support during tutor-student interactions, providing assistance on request. This feedback is concise, allowing tutors to quickly read and continue teaching. It includes two main components: (1) a description of students' current engagement and how the tutor engages them, and (2) teaching strategy recommendations to address immediate issues. To generate (1), the system uses the latest tutor-student conversation to summarize the current situation. By incorporating matched teaching strategies (Subsections \ref{subsec:survey2} and \ref{subsec:iteration}), it identifies effective strategies for the three students and provides actionable insights for the tutor.
Immediate Feedback allows tutors to create a practice loop by testing strategies, observing student responses, and adjusting their approach if needed. When a strategy isn’t effective, tutors can consult the feedback to find alternatives, expanding their repertoire and improving their application of strategies in real teaching scenarios.

\subsubsection{Asynchronous Feedback}
\label{subsec:Asynchronous Feedback }
Compared to Immediate Feedback, Asynchronous Feedback is more detailed and comprehensive, designed to help tutors reflect on their practice for improvement. To structure Asynchronous Feedback we use the Integrated Reflective Cycle~\cite{bassot2015reflective}, which includes four steps: (1) describing the experience, (2) reflecting on what went well and what could be improved, (3) connecting the experience to broader theories, and (4) using the reflection for future preparation. %Following this structure, Asynchronous Feedback is organized around four core components. 
Tab.~\ref{tab:asynchoronous_feedback} outlines the themes of each step in the context of Asynchronous Feedback and provides examples of feedback content.

\aptLtoX{\begin{table}
\centering
\caption{Asynchronous Feedback generated following the Integrated Reflective Cycle.}
\label{tab:asynchoronous_feedback}
\begin{tabular}{lp{80pt}p{110pt}}
\toprule
Stage                             & Explanation                                           & Feedback Example                                                                                                                                                                                                  \\
\midrule
Overview              & Dialogue Overview                                     & The
  conversation depicts a tutor facilitating a math problem-solving session
  with three students, Ethan, Chloe, and Noah. The tutor asks them to
  calculate.                                             \\
  \hline
Reflection &  Student Engagement Analysis                 & Ethan's Cognitive Engagement: 
Ethan shows high cognitive engagement by quickly providing answers and later explaining the math problem to his peers.                                                               \\
\hline
Theory                           & Evaluation of Tutor's Strategies                   & Distributed Questioning:
  Calling on each student to answer questions increased participation but
  highlighted varying levels of engagement and understanding.                                                  \\
  \hline
Preparation               & Suggestions and Recommendations for Future Interactions & Targeted Support for Noah: Provide Noah with more direct and supportive instructional strategies, such as one-on-one follow-ups or scaffolded questioning to build his confidence and understanding incrementally. \\
\bottomrule
\end{tabular}
\end{table}}{\begin{table}
\centering
\caption{Asynchronous Feedback generated following the Integrated Reflective Cycle.}
\label{tab:asynchoronous_feedback}
\begin{tblr}{
  width = \linewidth,
  colspec = {Q[16]Q[29]Q[55]},
  row{1} = {c},
  hlines,
}
Stage                             & Explanation                                           & Feedback Example                                                                                                                                                                                                  \\
Overview              & Dialogue Overview                                     & The
  conversation depicts a tutor facilitating a math problem-solving session
  with three students, Ethan, Chloe, and Noah. The tutor asks them to
  calculate.                                             \\
Reflection &  Student Engagement Analysis                 & Ethan's Cognitive Engagement: 
Ethan shows high cognitive engagement by quickly providing answers and later explaining the math problem to his peers.                                                               \\
Theory                           & Evaluation of Tutor's Strategies                   & Distributed Questioning:
  Calling on each student to answer questions increased participation but
  highlighted varying levels of engagement and understanding.                                                  \\
Preparation               & Suggestions and Recommendations for Future Interactions & Targeted Support for Noah: Provide Noah with more direct and supportive instructional strategies, such as one-on-one follow-ups or scaffolded questioning to build his confidence and understanding incrementally. 
\end{tblr}
\end{table}}

\section{User Study: Evaluation of TutorUp System}
\label{sec:user_study}

\textit{TutorUp} is designed to help novice tutors practice teaching strategies for promoting student engagement in online tutoring through scenario-based training. To validate its effectiveness, we employed mixed methods and designed a within-subjects study comparing \textit{TutorUp} with a baseline system lacking its core features--further introduced in Subsection~\ref{subsec:baseline}. We analyzed participant evaluations of both systems to highlight the advantages of \textit{TutorUp} in key areas. Additionally, an expert-driven qualitative assessment was conducted to evaluate tutor performance in addressing student engagement challenges after training with each system. 
% To assess the effectiveness of our proposed system, w

\subsection{Participants} 
\label{subsec:participants}

We recruited 16 participants (6 females and 10 males) proficient in English as the targeted users for training with \textit{TutorUp} and the baseline system. Among them, four were novice tutors from JANN, and twelve were full-time university students from China and the U.S. All participants were novice online tutors or interested in online teaching, with less than 30 hours of experience and no formal tutor training. Sessions were conducted via Zoom, and participants were compensated at a rate of \$10 per hour. 
For the qualitative assessment, we recruited two experienced online tutors, each with over 15 months of online teaching experience. They were tasked with evaluating the participants' test results using co-designed metrics (details in Subsection~\ref{subsec:qualitative_assessment})

\subsection{Baseline System}
\label{subsec:baseline}

\begin{figure*}
    \centering
    \includegraphics[width=\textwidth]{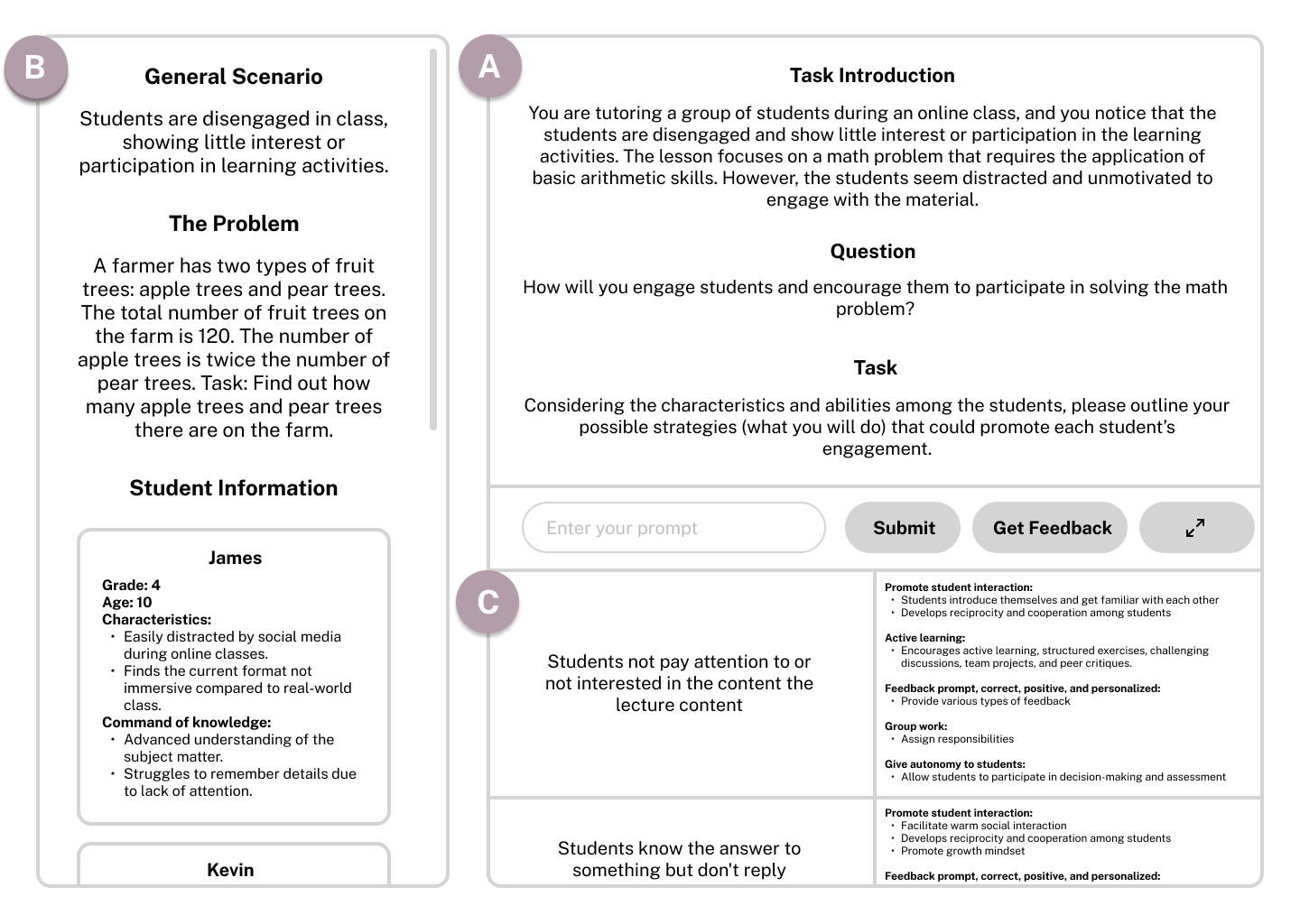}
    \caption{Baseline System: The baseline system consists of three main panels:  A) \textit{Task Introduction Panel: }This panel introduces the general scenario to the tutors, prompting them to consider how they would engage students in the given scenario. Tutors are then asked to write down their proposed strategies for student engagement.  B) \textit{Scenario Information Panel:} This panel provides the scenario's theme, the math problem, and relevant student information, all of which are also included in \textit{TutorUp}.  C) \textit{Feedback Panel:} After tutors complete their responses, this panel offers scenario-strategy pairs as feedback. These pairs are also used as part of the feedback prompts in \textit{TutorUp}.}
    \label{fig:baseline}
\end{figure*}

To test how our system performed versus other scenario-based tutor training tools, we built a baseline system. Specifically, we wanted to explore how the core features--the scenario scheme (LLMs simulating students) and the feedback scheme (LLMs providing personalized feedback based on learning theories)--performed in comparison to other systems. The design of the baseline system was inspired by the tutoring system created by \citet{personalizedlearning}, which also provides tutoring scenarios and feedback for tutors. However, the scenarios are described only in text and do not simulate a realistic multi-turn conversation and the feedback is not tailored to the tutor's responses~\cite{developmentofscenario, whenthetutorbecomes}. Specifically, while keeping other information and design elements (scenario theme, student information, math problems, and matched teaching strategies) unchanged, we removed the mechanism of simulating students to present the scenario. Instead, the scenario was presented through static text descriptions. We also removed the use of LLMs in providing feedback and instead directly provided a corresponding strategy-scenario matching table (Section~\ref{subsec:survey2}). The key differences between the baseline system and \textit{TutorUp} in scenario and feedback scheme are listed in Tab. \ref{baseline-differences}.

To receive training from the baseline system, users first review the task introduction panel (Fig. \ref{fig:baseline} A) to see the scenario description, which provides the context of the present student engagement challenges. Users are then prompted to think about how to engage these students. Without simulated students for them to interact with, their task is to write down teaching strategies they would implement to address the scenario, combined with the information on scenario background panel(Fig. \ref{fig:baseline} B) listed on the left. When users finish writing down teaching strategies, they can submit their responses and click ``Get Feedback'' to view a table (Fig. \ref{fig:baseline} C) of scenario-strategy pairs as feedback.

\subsection{Study Setup}
\label{subsec:setup}
To counterbalance potential order effects, the recruited novice tutors were divided into two groups with eight tutors each. We asked one group to experience the Baseline condition first, followed by the \textit{TutorUp} condition, while the other group followed the reverse order. Since both the baseline system and \textit{TutorUp} are scenario-based training systems, we randomized the four designed scenario assignments across participants to avoid bias and ensure a balanced training experience. To prevent cross-condition influence in this within-subjects study, different math problems were presented in each condition to maintain contextual distinction. 

To assess the training effectiveness, we designed a separate test for participants to complete after each system’s training. 
On one hand, the test should be realistic, so ideally, tutors would teach in actual online learning scenarios with disengaged students to evaluate their performance. On the other hand, setting up real scenarios is costly and could potentially have negative impacts on students. Therefore, our test uses simulated scenarios, with the same mechanism as \textit{TutorUp}, but without the reset or feedback functions. To minimize the potential effects caused by the test system, which differs from the baseline condition but is similar to the \textit{TutorUp} condition, users of the baseline system were asked to first learn and familiarize themselves with the test system in order to adapt to the simulated teaching scenario.

\aptLtoX{\begin{table}
\centering
\caption{Comparison between instructional conditions within the Baseline and the \textit{TutorUp} system. The Baseline system features textual scenario descriptions, a writing tasks, and summative feedback. The \textit{TutorUp} system implements scenario-based training via interactions with LLM-based conversational agents and LLM-based feedback on user inputs and teaching strategies.}
\label{baseline-differences}
\begin{tabular}{p{50pt}p{70pt}p{120pt}}
\toprule
Condition & Baseline & \textit{TutorUp} \\
\midrule
Scenario Scheme & Text description & Interactions with simulated students\\
\hline
Feedback Scheme & A list of strategies & LLM-generated immediate and asynchronous feedback \\
\hline
Training Task   & Write down the measures and strategies to engage students with strategy list & Have conversations with simulated students to engage them with immediate and asynchronous feedback \\
\hline
Test Task & Engage simulated students for the same scenario where they have been trained &                            \\
\bottomrule    
\end{tabular}
\end{table}}{\begin{table}
\centering
\caption{Comparison between instructional conditions within the Baseline and the \textit{TutorUp} system. The Baseline system features textual scenario descriptions, a writing tasks, and summative feedback. The \textit{TutorUp} system implements scenario-based training via interactions with LLM-based conversational agents and LLM-based feedback on user inputs and teaching strategies.}
\label{baseline-differences}
\begin{tblr}{
  width = 0.9\linewidth,
  colspec = {Q[20]Q[40]Q[40]},
  cells = {l},
  cell{5}{2} = {c=2}{0.72\linewidth},
  hlines,
  vline{2-3} = {1-4}{},
  vline{2} = {5}{},
}
Condition & Baseline & \textit{TutorUp} \\
Scenario Scheme & Text description & Interactions with simulated students\\
Feedback Scheme & A list of strategies & LLM-generated immediate and asynchronous feedback \\
Training Task   & Write down the measures and strategies to engage students with strategy list & Have conversations with simulated students to engage them with immediate and asynchronous feedback \\
Test Task & Engage simulated students for the same scenario where they have been trained &                                
\end{tblr}
\end{table}}

\subsection{Study Procedure}

The study was conducted online via Zoom and was recorded for review and verification. Each participant, based on their assigned group, received training with both systems in sequence.

In each system, participants received training on one of different designed scenarios. During the baseline system training, tutors stopped whenever they felt they had done everything they would typically do to engage students and then review the feedback. In the \textit{TutorUp} training, tutors could engage with simulated students and request immediate feedback at any time. They could reset the scenario and try multiple approaches, and once they felt they had received sufficient training, they could end the session and review the asynchronous feedback. After each training session, participants completed a test, where they had 10 minutes to engage these simulated students. Each test after the training featured the same scenario and students as in the training, but with a different math problem to ensure participants apply the skills they've learned, rather than relying on rote memorization of the scenario. In both systems, participants were asked to try to learn and remember something useful and helpful from the training and apply it during the simulated teaching test. 

After completing each system’s training, they took a separate test designed by us, resulting in two sets of test results. The training and test records from all participants were collected for a comprehensive comparative analysis. Following each training and test, participants completed a post-task survey consisting of a questionnaire with 5-point Likert scale questions derived from the existing literature~\cite{kirkpatrick1994, howcroft2020twenty}, as listed in Tab. \ref{tab:questionnaire}. 

After both training, we also conducted an interview with open-ended questions to gather participants' thoughts on the key features of \textit{TutorUp} in terms of Scenario Simulation, Feedback scheme, and Overall system usage.

\subsection{Qualitative Assessment}
\label{subsec:qualitative_assessment}

Two experienced tutors were asked to provide a qualitative assessment of participants' test results to determine which system led to better training outcomes. The test results consisted of dialogue records from tutoring sessions with the simulated students. In collaboration with the tutors, we designed four evaluation criteria (Fig. ~\ref{fig:expert_ratings}). Each dialogue record was rated on a scale of 1 to 3. To ensure fairness in the assessment, the tutors were blinded to whether each dialogue came from the \textit{TutorUp} system or the baseline system.

\aptLtoX{\begin{table}
\centering
\caption{Our questionnaire focuses on four central aspects: Relevance and Motivation (Q1-Q2), System Effectiveness (Q3-Q4), Skill Improvement and Confidence (Q5-Q7), and System Usability (Q9-Q11).}
\label{tab:questionnaire} % Label for referencing the table
\begin{tabular}{p{60pt}lp{170pt}}
\toprule
Category                         && Question~                                                                                                     \\
\midrule
Relevance and Motivation         & Q1:       & You find the training relevant to your jobs                                                        \\
                                 & Q2:       & You find yourself prompted to enhance your abilities to engage students                            \\
                                 \hline
System Effectiveness             & Q3        & You find the class scenario scheme effective and practical                                         \\
                                 & Q4        & You find the feedback provided during the training helpful and constructive                        \\
                                 \hline
Skill Improvement and Confidence & Q5        & You believe such training will be beneficial to students' engagement problems when teaching online \\
                                 & Q6        & You acquire useful strategies for managing students' engagement problems                           \\
                                 & Q7        & You feel more confident about yourself for dealing with students' engagement problems               \\
                                 & Q8        & You will apply what you learned during training when you are back on the job                       \\
                                 \hline
System Usability and Engagement  & Q9        & You find the system easy to use                                                                    \\
                                 & Q10       & You anticipate using this system frequently.                                                       \\
                                 & Q11       & You find the training engaging                                                                     \\
                                 \bottomrule
\end{tabular}
\end{table}}{\begin{table}
\centering
\caption{Our questionnaire focuses on four central aspects: Relevance and Motivation (Q1-Q2), System Effectiveness (Q3-Q4), Skill Improvement and Confidence (Q5-Q7), and System Usability (Q9-Q11).}
\label{tab:questionnaire} % Label for referencing the table

\begin{tblr}{
  width = \linewidth,
  colspec = {Q[238]Q[46]Q[654]},
  row{1} = {c},
  cell{1}{2} = {c=2}{0.7\linewidth},
  cell{2}{1} = {r=2}{c},
  cell{2}{2} = {c},
  cell{3}{2} = {c},
  cell{4}{1} = {r=2}{c},
  cell{4}{2} = {c},
  cell{5}{2} = {c},
  cell{6}{1} = {r=4}{c},
  cell{6}{2} = {c},
  cell{7}{2} = {c},
  cell{8}{2} = {c},
  cell{9}{2} = {c},
  cell{10}{1} = {r=3}{c},
  cell{10}{2} = {c},
  cell{11}{2} = {c},
  cell{12}{2} = {c},
  hline{1-2,4,6,10,13} = {-}{},
}
Category                         & Question~ &                                                                                                    \\
Relevance and Motivation         & Q1:       & You find the training relevant to your jobs                                                        \\
                                 & Q2:       & You find yourself prompted to enhance your abilities to engage students                            \\
System Effectiveness             & Q3        & You find the class scenario scheme effective and practical                                         \\
                                 & Q4        & You find the feedback provided during the training helpful and constructive                        \\
Skill Improvement and Confidence & Q5        & You believe such training will be beneficial to students' engagement problems when teaching online \\
                                 & Q6        & You acquire useful strategies for managing students' engagement problems                           \\
                                 & Q7        & You feel more confident about yourself for dealing with students' engagement problems               \\
                                 & Q8        & You will apply what you learned during training when you are back on the job                       \\
System Usability and Engagement  & Q9        & You find the system easy to use                                                                    \\
                                 & Q10       & You anticipate using this system frequently.                                                       \\
                                 & Q11       & You find the training engaging                                                                     
\end{tblr}
\end{table}}

\subsection{Hypothesis}
We propose the following alternative hypotheses, informed by prior literature on tutor training systems ~\cite{kirkpatrick1994, howcroft2020twenty}, targeting both participants' evaluations of the system and the qualitative assessment from the experienced tutors.

H1. \textit{TutorUp} performs better than the baseline system in terms of training relevance (H1a) and  necessity (H1b)

H2. \textit{TutorUp} performs better than the baseline system in terms of training effectiveness. Specifically, \textit{TutorUp} features more practical scenario presentation (H2a) and usefulness of feedback (H2b) compared to the baseline system.

H3. \textit{TutorUp} performs better than the baseline system in terms of tutors' skill and confidence increase. Specifically, \textit{TutorUp} achieves higher increase in confidence (H3a), applicability (H3b), strategies (H3c), and future benefits (H3d) to tackle engagement challenges.

H4. \textit{TutorUp} performs better than the baseline system in terms of usability. Specifically, the system is easier to use (H4a), more engaging and enjoyable (H4b), and more likely to be used in the future (H4c).

H5. Tutors trained by \textit{TutorUp} use strategies more appropriately compared to the baseline system. 

H6. \textit{TutorUp} encourages higher student engagement compared to the baseline system. 
% Specifically, students are more prompted to be actively engaged in learning through \textit{TutorUp}.

H7. \textit{TutorUp} enables tutors to deliver strategies in a better manner compared to the baseline system. Specifically, tutors' instructions to apply strategies are more accessible (H7a) and effective (H7b) for students to engage. 
\section{Study Results}
\begin{figure}[h]
    \centering
    \begin{subfigure}[t]{0.50\textwidth}
        \centering
        \includegraphics[width=\textwidth]{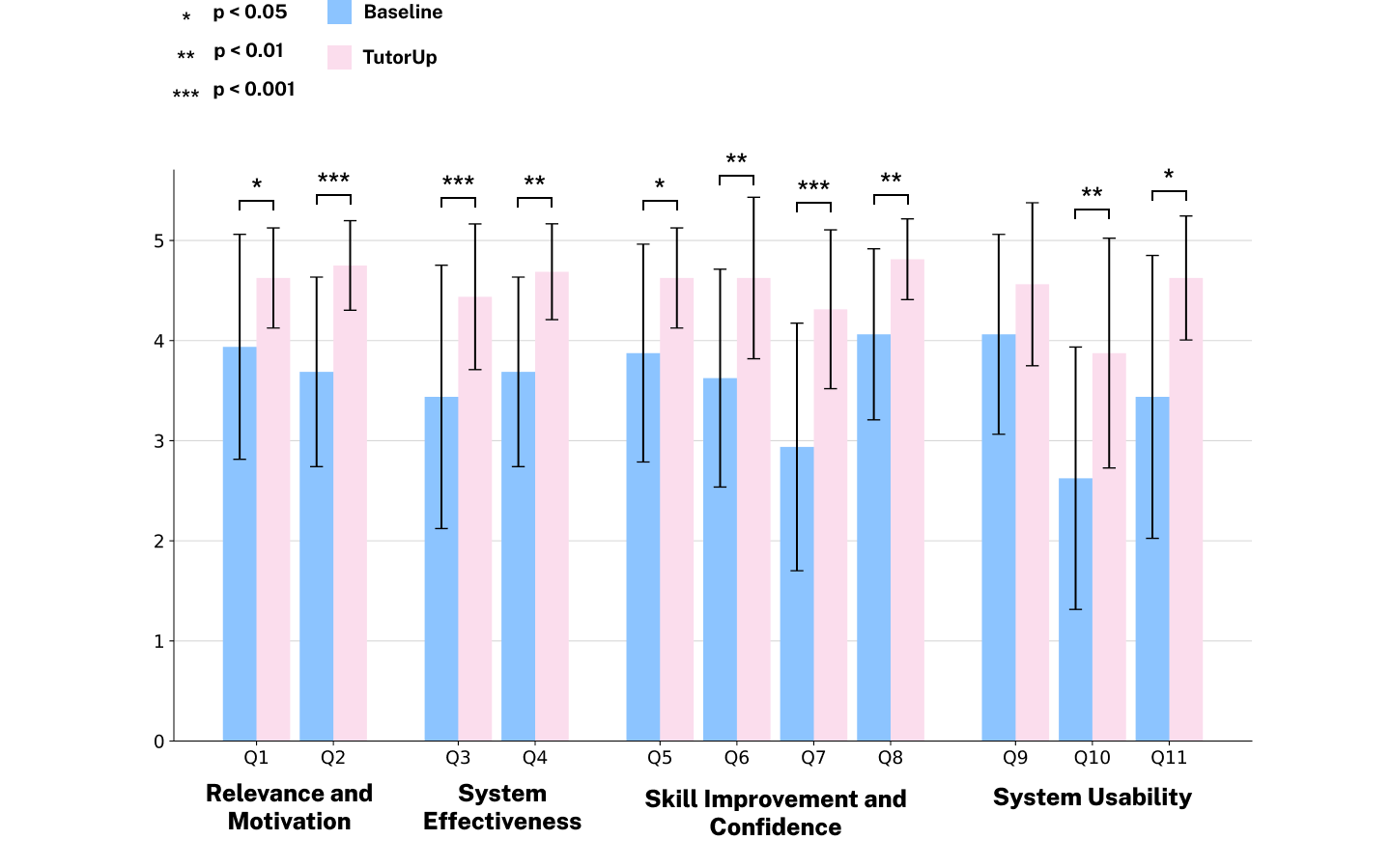}
        \caption{Participants' Rating}
        \label{fig:participant_ratings}
    \end{subfigure}%
    \hfill
    \begin{subfigure}[t]{0.50\textwidth}
        \centering
        \includegraphics[width=\textwidth]{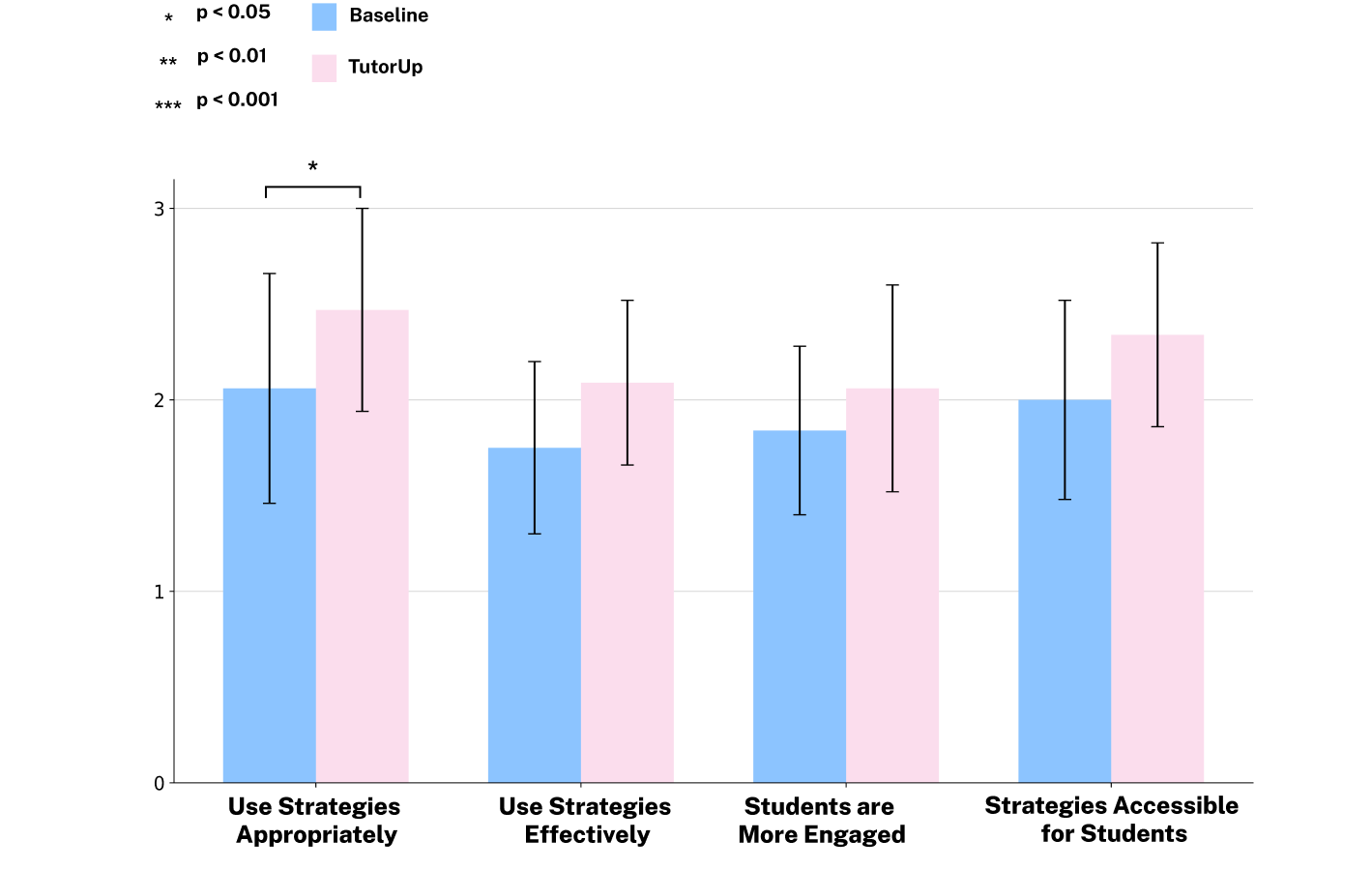}
        \caption{Qualitative Assessment}
        \label{fig:expert_ratings}
    \end{subfigure}
    \caption{(a): Means and standard errors of baseline system and TutorUp
on a 5-point Likert measured by User-Study participants; (b) Means and standard errors of baseline system and \textit{TutorUp}
on test results on a 3-point Likert measured by experts.}
    \label{fig:combined_figures}
\end{figure}
We present evaluation results for the Survey, Test Assessment, and Interview. The full $p$-value data for the first two are in Appendix \ref{apdx:evaluation}.
\subsection{Survey Results}

We conducted a binomial test (equivalent to a Sign Test) to evaluate whether \textit{TutorUp} significantly enhances the training experience for novice tutors compared to the baseline system. Using a significance threshold of $p<0.05$, the results show that participants rated \textit{TutorUp} significantly higher across key areas, as illustrated in Fig. ~\ref{fig:expert_ratings}. Overall, all hypotheses were supported, except H4a, with $p$-values larger than $0.05$ in the binomial test. However, there was no significant difference ($p=0.90$) between \textit{TutorUp} and the baseline system in terms of ease of use, failing to support hypothesis H4a.  The reason might be that the layout of information presented in both the baseline and \textit{TutorUp} systems is quite similar. Based on the mean scores (Baseline Mean = $4.06$, \textit{TutorUp} Mean = $4.56$), both systems are generally easy to use, resulting in minimal differences.

\subsection{Test Assessment Results}

We averaged the test evaluations from two experienced tutors. Since the data from the four evaluation metrics followed a normal distribution, we performed a Paired Samples t-test on the tutors' test scores, with results shown in Fig. \ref{fig:expert_ratings}. Using a threshold of $p < 0.05$, we observed that participants performed significantly better in the \textit{Use Strategies Appropriately} aspect after completing the \textit{TutorUp} training (Mean = $2.47$, SD = $0.53$), compared to the baseline training (Mean = $2.06$, SD = $0.60$) ($p < 0.05$, supporting H5). However, for the \textit{Use Strategies Effectively} ($p = 0.09$), \textit{Students are More Engaged} ($p = 0.23$), and \textit{Strategies Accessible for Students} ($p = 0.10$) aspects, although the mean scores for \textit{TutorUp} were higher than for the baseline, the differences were not statistically significant (H6, H7a, H7b not supported). Due to time limitations, we only asked tutors to practice one scenario for each system, which may explain why the training effectiveness for strategy application is not obvious (H7a and H7b). The failure to support H6 is discussed in Section \ref{subsec:simulateEngagement}.

\subsection{Interview Results}
We summarized the ideas conveyed during the interview for the Scenario Simulation, Feedback scheme, and Overall system usage.
\subsubsection{Scenario Simulation}
We asked tutors about their opinions on scenario simulation in three perspectives: (1) scenario theme, (2) student information and student simulation, and (3) math problem. \textbf{\textit{Scenario theme: }} 
Several participants (P6, P9, P12, P14) expressed the importance and clarity of the scenario theme description. P6 noted that it is concise and clear, providing a detailed account of the issues and student context. However, some participants (P4, P7, P11) found the scenario theme descriptions useful but redundant due to overlap with the student information. \textbf{\textit{Student information:}} All participants expressed that the detailed student information was very helpful for applying strategies to specific scenarios. Most of them felt that the student simulations were quite realistic, noting that the students’ tone and dialogue aligned well with their character profiles. However, participants also pointed out certain aspects of the student design that felt unrealistic. P7 mentioned that the language used by the students did not match their age (around 10 years old), as it's unlikely for children that age to speak without grammatical errors. P2 commented that, in their home country (Mexico), children around that age are not allowed to use computers, which made the scenario feel unrealistic to them. Additionally, P9 felt that the students’ behavior was too straightforward and logical, lacking the complexity expected of real children, while P10 found the students to be too ``slow'' as they struggled with simple math problems that should have been easy for them. \textbf{\textit{Math problem:}} Several participants (P6, P8, P10, P13, P15) agreed that using math problems as a context is beneficial because they involve clear steps and solutions, making them more logically structured compared to subjects like literature. This helps users focus on the issue of student engagement. However, some participants (P3, P7, P8, P15) felt that the current problems in the system were too simple. P14 suggested incorporating other types of questions such as open-ended ones for greater variety.

\subsubsection{Immediate and Asynchronous Feedback}
\label{sucsec:immediate}

All participants expressed their overall appreciation for the feedback scheme in the \textit{TutorUp} system, describing it as personalized and useful. 
Regarding \textbf{\textit{immediate feedback}}, participants emphasized their preference for its real-time nature and practical applicability. As P13 stated, ``Immediate feedback provides timely suggestions that I can apply right away, creating a positive loop between theory and practice.'' P7 also noted that ``the immediate feedback helps connect the strategy to the current scenario, which aids memory and application.'' However, P6 mentioned that the immediate feedback seems to focus on the most recent dialog round, suggesting that covering a broader range of interactions might be beneficial. 
Regarding \textbf{\textit{asynchronous feedback}}, many participants (P1–P10) expressed an appreciation for its completeness and richness, finding it useful and comprehensive. However, some participants (P12, P16) felt it was too lengthy and suggested it should focus more on practical advice. 
Regarding \textbf{\textit{both}} types of feedback, P4 and P11 mentioned that instead of just being told what strategies to try, they would prefer more detailed guidance on how to implement those strategies effectively. Although many participants expressed that the strategies in the feedback were very practical, some mentioned that after applying these strategies as suggested by the system, the simulated students did not appear to become more engaged.

\subsubsection{System Usage and Improvements}
Most participants agreed that \textit{TutorUp} is straightforward, clear, and easy to use. P13 specifically described it as ``overall convenient, intuitive, with a clear layout.'' Participants also offered numerous suggestions for improving key system features. A major area for improvement, mentioned by P2, P3, P5, and P13, was adding more diverse scenarios and varied scenario designs. Some also suggested enhancing the system’s customization options, such as allowing users to edit math problems (P10) and customize the personalities of simulated students and scenarios (P3 and P5). Additionally, participants hoped for more visual interactivity, like adding student avatars and incorporating embedded links or slides into the system, as suggested by P11.

\section{Discussion}

\subsection{Enhancing Novice Teachers' Understanding of Student Disengagement}
\label{subsec:enhancing_understanding}

In this work, we summarized scenarios of online tutoring challenges through survey questions like, ``Have you employed a specific strategy and how and when did you use it?'' The scenarios were based on tutors' descriptions of student situations (e.g., fast learners get bored easily). However, only a small number of responses ($138/1326$) provided detailed scenario descriptions, with many answers being vague, such as ``always'' or ``at the beginning.'' This suggests that while tutors can sense disengagement, they struggle to accurately link it to students' underlying states behaviors. While other studies~\cite{fredricks2004school,joshi2022behavioral} on engagement focus on abstract dimensions like cognitive, emotional, and social engagement, our study uses more observable and concrete features, such as a student’s behavior, knowledge level, and initial actions, to represent engagement. This shift to tangible metrics makes signs of disengagement more visible and helps tutors better understand student behavior.

\subsection{Interactive Scenarios Foster Immersive Learning and Adaptive Strategy Development}

Our study found that by providing more interactive and concrete student simulations, \textit{TutorUp} enables tutors to engage directly with specific problems and apply strategies more practically and effectively. The dialogue-based format fosters a relaxed, realistic environment, encouraging tutors to experiment with different strategies, thus reducing the fear and anxiety of making mistakes that they might face in real classrooms.
Additionally, in our system, students' reactions serve as another form of feedback, closely mirroring real-life scenarios. Teachers can assess whether their chosen strategies are effective based on the simulated students' responses and adjust their approach accordingly. The presence of a reset feature encourages teachers to undo actions and try different strategies, providing additional practice opportunities. In this way, the reactive scenario-based training system demonstrates its distinct advantages.

\subsection{Simulating Natural Online Students-Tutor Session }

Inspired by prior work on LLMs for creating and extending logically coherent stories and scripts ~\cite{mirowski2023co,shakeri2021saga,grigis2024playwriting}, we introduced the BigPicture Agent to manage the sequence of interactions between students and tutors. It handled student responses, ensuring logical progression in the dialogue. For example, if a tutor asks a question, the BigPicture Agent ensures the correct student responds, maintaining conversational consistency.

During development, we found it necessary to implement some rules for the BigPicture agent to structure dialogues. For instance, when the tutor greets the students, all students must respond to create a polite and friendly environment. Additionally, consecutive student responses were limited to leave space for the tutor’s instructions. However, in practice, as the conversation lengthened, LLM-based agents sometimes deviated from instructions, leading to inconsistent behavior, such as students ``forgetting'' previous parts of the conversation or responding only one at a time unless directly addressed by the tutor. This highlights the limitations of current LLMs in simulating dialogue, as they cannot perfectly replicate the dynamics of real-world interactions. Future work could focus on enhancing memory management and context retention in LLMs to handle long, dynamic conversations more effectively, possibly by incorporating techniques like context tracking or memory augmentation, which have been explored in recent research~\cite{wang2024augmenting, yi2024survey}. Additionally, hallucinations can occur, where the model fabricates unrealistic or inconsistent student behaviors, like a disengaged student suddenly giving detailed answers. To mitigate this, we designed student prompts to enforce consistency and limit unrealistic responses. For example, the prompt explicitly instructs that the student behavior should be gradually adjusted based on tutor interactions, but not instantly, which helps prevent the model from ``hallucinating'' a sudden shift in behavior. Future research will focus on developing more accurate student simulations and develop quantitative measures to evaluate their accuracy across various dimensions (e.g., cognitive and affective)~\cite{kaser2024simulated}.

\subsection{Using LLMs to Simulate Disengaged Students}
\label{subsec:simulateEngagement}

In our student agent's prompt, we did not specify how the students should \textit{change} their engagement levels over time in response to the tutors' teaching. Instead, we provided basic personality traits and instructed the agents to follow these traits. Following initial disengaged behaviors, their engagement would change based on the tutor's instructions. In other words, when and how much the students' engagement shifts, as well as how that change is displayed, is entirely determined by the LLM. This approach explores how LLMs model human-like behavior, but it also explains why H6, which supposes \textit{TutorUp} encourages higher student engagement compared to the baseline system, does not hold in the User Study. Despite the tutor using more appropriate strategies, the LLMs simulation may not reflect a consistent shift in student engagement as intended, as the model decides when and how engagement changes. Related work~\cite{shu2023you} has noted that current LLMs are unable to consistently and accurately model the subtle cognitive and psychological behaviors of humans. Nevertheless, even if the simulations are not perfectly accurate, tutors can still familiarize themselves with the teaching strategies and prototypical disengagement patterns, which can help improve their responses and adaptability in real-life situations.

%%%%%%%%%%%%%%%%%%%%%%%%%%%%%%%%%%%%%%%%%%%%%%%%%%%%%%%%%%%%
\section{Limitations and Future Work}

\subsection{The Presentation and Measurement of Engagement Status}
Since the students are simulated, and cannot accurately mimic the true state of human students, using precise quantitative data for analysis does not seem meaningful. So we utilize the LLM to give analysis of changes in student engagement (Subsection~\ref{subsec:Asynchronous Feedback }), considering it qualitatively through cognitive, behavioral, emotional, and collaborative aspects ~\cite{joshi2022behavioral}, over tutor-students dialogues. This qualitative assessment on engagement fails to offer tutors precise numerical metrics. Additionally, we allowed the LLMs to fully control how student engagement changes over time, which is likely not fully representative of human students. In future work, we plan to explore better methods for quantifying engagement to help tutors more accurately assess whether their strategies are effective and whether student engagement has truly shifted. We also aim to improve the prompts to make the LLMs’ simulation of student engagement more natural and realistic.

\subsection{More Fine-grained Scenario Design}

% We identified student disengagement scenarios from survey responses and designed student personas based on them. However, a
As discussed in Subsection ~\ref{subsec:enhancing_understanding}, novice tutors often struggle to accurately perceive disengagement, resulting in survey results that lacked detailed descriptions of what disengaged students look like. Instead, tutors provided broad characteristics limiting the accuracy and realism of our student personas.

Another limitation is that we did not consider factors such as gender, race, or cultural background, which can influence student behavior, learning abilities and communication styles~\cite{lee23generative}. For example, P3 in our user study mentioned that students aged 10-11 in their country would neither have internet access nor express themselves with such logical precision, making the scenario unrealistic for them.
Thus, while our current approach represents a general attempt at student design, future work should consider more diverse factors and ensure scenarios are authentic across different cultural contexts. Future iterations could also include greater customization options, allowing tutors to extend TutorUp to other domains, such as science, language arts, or social studies, by tailoring content to each domain's learning objectives. Moreover, the system’s conversational and scenario-based features could also be adapted to other settings, such as clinical environments for medical training or sales training for practicing customer interactions.

% \subsection{More Reasonable and Comprehensive Depiction of Scenarios}
\subsection{More Realistic Scenarios Depiction}

We used LLMs to simulate disengaged students through prompt design. We found that the student agents generally followed their assigned personalities and adjusted their engagement based on the tutor’s instructions, though not always perfectly. Participants (P7 and P8) noted that the simulated students were ``too smart'' and ``lacked complexity.'' This highlights one limitation of using LLMs to simulate young, disengaged students, as they sometimes fail to mimic the illogical or unpredictable behaviors typical of this age group, especially when simulating inattentiveness or poor performance. This "overly smart" behavior is a key challenge in using LLMs to model younger students accurately~\cite{gpteach,argyle2023out}. To address these limitations, future work should focus on improving the realism of LLM-based student simulations. This could involve refining prompt designs to capture the unpredictable behaviors of disengaged or struggling students, as well as using fine-tuning techniques with age-specific data. Introducing randomness in student behaviors could also enhance the realism of the interactions. 

Additionally, establishing guidance for measuring realism could be valuable for future studies. In our paper, we evaluated realism by asking users to compare the results of simulations with those observed in the real world, which is subjective and based on impression. One solution to solve this limitation is to calculate the similarity metrics (e.g.,cosine similarity and Jaccard similarity) between the simulated dialogue and real student-tutor conversations, which can assess how closely the dialogues and interaction patterns of the simulated students reflect real-world scenarios~\cite{lau2016empirical, niwattanakul2013using}.

\label{sec:discussion}

\section{Conclusion}
% This work presents TutorUp, an scenario-based tutor training system that focused on student online engagement problems. It utilized LLMs to present scenarios by simulating disengaged students for tutor to practise teaching. To provide an effective training, we introduced 1) reactive disengaged scenarios, 2) BigPicture-Character prompting pipeline and 3) immediate and asynchronous feedback scheme to facilitate tutor training process. Our user study with 16 novice tutors showed that TutorUp provides an effective training for novice tutors to deal with students online engagement problems, with useful strategies learnt and practiced.  We anticipate that our work can provide experiences on creating simulated scenario-based training systems and support for using LLMs to simulate population with  specific features. 

This work introduced \textit{TutorUp}, a scenario-based tutor training system designed to address student engagement challenges in online learning environments. \textit{TutorUp} leverages large language models (LLMs) to simulate disengaged students, allowing tutors to practice teaching strategies in a safe environment that mimics real teaching scenarios. To enhance training effectiveness, we implemented three key components: (1) reactive disengagement scenarios, (2) a BigPicture-Character prompting pipeline, and (3) a system that provides immediate and asynchronous feedback to support the training process.
Our user study with 16 novice tutors demonstrated that \textit{TutorUp} effectively equips tutors with practical strategies for managing student engagement challenges in online settings. Tutors were able to learn and apply useful strategies through the system. We hope that our work will contribute valuable insights for the development of scenario-based training systems and the utility of LLMs to simulate populations with specific characteristics.
\label{sec:conclusion}

\bibliographystyle{ACM-Reference-Format}
\bibliography{bibliography}

\newpage
\appendix

\section{Ten Strategies Summarized from Literature Review }
\label{apdx:strategies}
Strategies with multiple instances are input to prompt Immediate Feedback and Asynchronous Feedback in \textit{TutorUp} and serve as feedback for Baseline System.

\aptLtoX{\begin{table}[H]
\centering
\caption{Strategies summarized from Literature Review with two instances for each.}
\begin{tabular}{p{125pt}p{125pt}}
\toprule
\textbf{Strategy Category}                                 & \textbf{Strategy Instance}                                            \\
\midrule
Show empathy and respect toward students                   & Refer to students by name~                                            \\
                                                           & Demonstrate concern for their student ~                               \\
                                                           \hline
Promote peer interaction                                   & Encourages contact between students and faculty~                      \\
                                                           & Develops reciprocity and cooperation among students.~                 \\
                                                           \hline
Give prompt, correct, positive, and personalized Feedback~ & Encouragement to positive behavior ~                                  \\
                                                           & Be Free with Praise and Constructive in Criticism~                    \\
                                                           \hline
Promote persistence                                        & Don‘t allow students to give up, scaffold until they succeed ~        \\
\hline
                                                           & Treat each student as capable - believe that students want to learn ~ \\
                                                           \hline
Maintain active learning                                   & Encourage challenging discussions, team projects, and peer critiques  \\
\hline
                                                           & Expect active participation ~                                         \\
Set Clear goals~                                           & Show the Need for the Lesson~                                         \\
                                                           & Group work                                                            \\
                                                           \hline
Give autonomy to students                                  & Allow students to participate in decision-making and assessment ~     \\
\hline
                                                           & Take feedback from students~                                          \\
Promote group work                                         & Assign Responsibilities~                                              \\
                                                           & Ask students to answer each others questions ~                        \\
                                                           \hline
Set time constraint                                        & Emphasizes time on task.~                                             \\
                                                           & Ask for how many time students need                                   \\
                                                           \hline
Set behavioral expectations                                & Set clear expectation on student behavior                             \\
                                                           & Insist that students show respect and care for one another~~          \\
                                                           \bottomrule
\end{tabular}
\end{table}}{\begin{table}[H]
\centering
\caption{Strategies summarized from Literature Review with two instances for each.}
\begin{tblr}{
  width = \linewidth,
  colspec = {Q[431]Q[506]},
  cell{2}{1} = {r=2}{},
  cell{4}{1} = {r=2}{},
  cell{6}{1} = {r=2}{},
  cell{8}{1} = {r=2}{},
  cell{10}{1} = {r=2}{},
  cell{12}{1} = {r=2}{},
  cell{14}{1} = {r=2}{},
  cell{16}{1} = {r=2}{},
  cell{18}{1} = {r=2}{},
  cell{20}{1} = {r=2}{},
  hline{1-2,4,6,8,10,12,14,16,18,20,22} = {-}{},
}
\textbf{Strategy Category}                                 & \textbf{Strategy Instance}                                            \\
Show empathy and respect toward students                   & Refer to students by name~                                            \\
                                                           & Demonstrate concern for their student ~                               \\
Promote peer interaction                                   & Encourages contact between students and faculty~                      \\
                                                           & Develops reciprocity and cooperation among students.~                 \\
Give prompt, correct, positive, and personalized Feedback~ & Encouragement to positive behavior ~                                  \\
                                                           & Be Free with Praise and Constructive in Criticism~                    \\
Promote persistence                                        & Don‘t allow students to give up, scaffold until they succeed ~        \\
                                                           & Treat each student as capable - believe that students want to learn ~ \\
Maintain active learning                                   & Encourage challenging discussions, team projects, and peer critiques  \\
                                                           & Expect active participation ~                                         \\
Set Clear goals~                                           & Show the Need for the Lesson~                                         \\
                                                           & Group work                                                            \\
Give autonomy to students                                  & Allow students to participate in decision-making and assessment ~     \\
                                                           & Take feedback from students~                                          \\
Promote group work                                         & Assign Responsibilities~                                              \\
                                                           & Ask students to answer each others questions ~                        \\
Set time constraint                                        & Emphasizes time on task.~                                             \\
                                                           & Ask for how many time students need                                   \\
Set behavioral expectations                                & Set clear expectation on student behavior                             \\
                                                           & Insist that students show respect and care for one another~~          
\end{tblr}
\end{table}}

% Here we have space to include additional information and materials. For example survey questions and detailed descriptions of the individual prompts used by the system.

\section{Evaluation results in tabular form}
\label{apdx:evaluation}
% \subsection{Evaluation results in tabular form}

These tables contain the same information as the bar charts in Fig. \ref{fig:participant_ratings} and Fig. \ref{fig:expert_ratings}.
% \label{fig:participant_ratings}
% \label{fig:expert_ratings}

% \usepackage{tabularray}
\aptLtoX{\begin{table}[H]
\centering
\caption{Statistics on Qualitative Assessment.}
\label{tab:qualitativeResult}
\begin{tabular}{llp{40pt}p{40pt}p{40pt}p{40pt}}
\toprule
Measurement           &      & \textbf{Use Strategies Appropriately} & \textbf{Use Strategies Effectively} & \textbf{Students are More Engaged.} & \textbf{Strategies Accessible for Students} \\
\midrule
Baseline              & Mean & 2.06                               & 1.75                                & 1.84                                & 2.00                                        \\
                      & SD   & 0.60                               & 0.45                                & 0.44                                & 0.52                                        \\
                      \hline
TutorUp               & Mean & 2.47                               & 2.09                                & 2.06                                & 2.34                                        \\
                      & SD   & 0.53                               & 0.43                                & 0.54                                & 0.48                                        \\
                      \hline
Shapiro-Wilk          &      & 0.05                               & 0.34                                & 0.12                                & 0.65                                        \\
\hline
Paired Samples t Test &      & 0.03                               & 0.09                                & 0.23                                & 0.10  \\
\bottomrule                                      
\end{tabular}
\end{table}}{\begin{table}[H]
\centering
\caption{Statistics on Qualitative Assessment.}
\label{tab:qualitativeResult}
\begin{tblr}{
  width = \linewidth,
  colspec = {Q[92]Q[71]Q[181]Q[173]Q[190]Q[225]},
  cells = {c},
  cell{1}{1} = {c=2}{0.162\linewidth},
  cell{2}{1} = {r=2}{},
  cell{4}{1} = {r=2}{},
  cell{6}{1} = {c=2}{0.162\linewidth},
  cell{7}{1} = {c=2}{0.162\linewidth},
  hline{1-2,4,6-8} = {-}{},
}
Measurement           &      & \textbf{Use Strategies Appropriately} & \textbf{Use Strategies Effectively} & \textbf{Students are More Engaged.} & \textbf{Strategies Accessible for Students} \\
Baseline              & Mean & 2.06                               & 1.75                                & 1.84                                & 2.00                                        \\
                      & SD   & 0.60                               & 0.45                                & 0.44                                & 0.52                                        \\
TutorUp               & Mean & 2.47                               & 2.09                                & 2.06                                & 2.34                                        \\
                      & SD   & 0.53                               & 0.43                                & 0.54                                & 0.48                                        \\
Shapiro-Wilk          &      & 0.05                               & 0.34                                & 0.12                                & 0.65                                        \\
Paired Samples t Test &      & 0.03                               & 0.09                                & 0.23                                & 0.10                                        
\end{tblr}
\end{table}}

\aptLtoX{\begin{table}[!h]
\centering
\caption{Statistics on User Study Result.}
\label{tab:userstudyResult}
\begin{tabular}{llp{30pt}p{30pt}p{30pt}p{30pt}p{30pt}p{30pt}p{30pt}p{30pt}p{30pt}p{30pt}p{30pt}}
\toprule
Aspect        &      & Relevance and Motivation &       & System Effectiveness &       & Skill Improvement and Confidence &       &       &       & System Usability &       &       \\
\midrule
Question      &      & Q1                       & Q2    & Q3                   & Q4    & Q5                               & Q6    & Q7    & Q8    & Q9               & Q10   & Q11   \\
\hline
Baseline      & Mean & 3.94                     & 3.69  & 3.44                 & 3.69  & 3.88                             & 3.63  & 2.94  & 4.06  & 4.06             & 2.63  & 3.94  \\
              & SD   & 1.12                     & 0.95  & 1.31                 & 0.95  & 1.09                             & 1.09  & 1.24  & 0.85  & 1.00             & 1.31  & 1.31  \\
              \hline
TutorUp       & Mean & 4.63                     & 4.75  & 4.44                 & 4.69  & 4.63                             & 4.63  & 4.31  & 4.81  & 4.56             & 3.88  & 4.63  \\
              & SD   & 0.50                     & 0.45  & 0.73                 & 0.48  & 0.50                             & 0.81  & 0.79  & 0.40  & 0.81             & 1.15  & 1.15  \\
              \hline
Binomial Test &      & .035                    & .000 & .001                & .001 & .020                            & .003 & .000 & .004 & .089            & .002 & .010 \\
\bottomrule
\end{tabular}
\end{table}}{\begin{table}[!h]
\centering
\caption{Statistics on User Study Result.}
\label{tab:userstudyResult}
\begin{tblr}{
  width = \linewidth,
  colspec = {Q[67]Q[56]Q[92]Q[92]Q[79]Q[79]Q[65]Q[65]Q[65]Q[65]Q[56]Q[56]Q[54]},
  cells = {c},
  cell{1}{1} = {c=2}{0.123\linewidth},
  cell{1}{3} = {c=2}{0.184\linewidth},
  cell{1}{5} = {c=2}{0.158\linewidth},
  cell{1}{7} = {c=4}{0.26\linewidth},
  cell{1}{11} = {c=3}{0.166\linewidth},
  cell{2}{1} = {c=2}{0.123\linewidth},
  cell{3}{1} = {r=2}{},
  cell{5}{1} = {r=2}{},
  cell{7}{1} = {c=2}{0.123\linewidth},
  hline{1-3,5,7-8} = {-}{},
}
Aspect        &      & Relevance and Motivation &       & System Effectiveness &       & Skill Improvement and Confidence &       &       &       & System Usability &       &       \\
Question      &      & Q1                       & Q2    & Q3                   & Q4    & Q5                               & Q6    & Q7    & Q8    & Q9               & Q10   & Q11   \\
Baseline      & Mean & 3.94                     & 3.69  & 3.44                 & 3.69  & 3.88                             & 3.63  & 2.94  & 4.06  & 4.06             & 2.63  & 3.94  \\
              & SD   & 1.12                     & 0.95  & 1.31                 & 0.95  & 1.09                             & 1.09  & 1.24  & 0.85  & 1.00             & 1.31  & 1.31  \\
TutorUp       & Mean & 4.63                     & 4.75  & 4.44                 & 4.69  & 4.63                             & 4.63  & 4.31  & 4.81  & 4.56             & 3.88  & 4.63  \\
              & SD   & 0.50                     & 0.45  & 0.73                 & 0.48  & 0.50                             & 0.81  & 0.79  & 0.40  & 0.81             & 1.15  & 1.15  \\
Binomial Test &      & .035                    & .000 & .001                & .001 & .020                            & .003 & .000 & .004 & .089            & .002 & .010 
\end{tblr}
\end{table}}

\section{Survey Questions}
\label{apdx:sureveyquestions}
\subsection{First Survey Questions}
\begin{enumerate}
    \item What difficulties or challenging situations did you face when you started giving the tutor's?
    \item What difficulties and challenging situations do you currently face during the tutor’s?

\end{enumerate}

\subsection{Second Survey Questions}

\textbf{Open-ended Questions: }

\begin{enumerate}
    \item What types of student engagement problems have you encountered?
    \item Why do you think students are disengaged?
    \item Which strategies have you used to increase students' engagement and when did you use them? Please list all strategies and the situations when you use them:
\end{enumerate}

\textbf{Strategy-specific Questions: }

\begin{enumerate}
    \item Did you show empathy and respect towards your students to improve students' engagement (e.g., referring to students by their names or demonstrating concerns for their well being)? How and when?
    \item Did you promote social community and belonging among your students to improve students' engagement (e.g., let students introduce themselves or facilitate warm social interaction between students)? How and when?
    \item Did you discuss behavioral expectations with your students to improve students' engagement (e.g., don't interrupt each other when talking or start tutoring sessions on time)? How and when?
    \item Did you discuss academic goals with your students to improve students' engagement? (e.g., master certain subject specific skills or personal long term goals that students want to achieve (go to university))? How and when?
    \item Did you promote active learning in your tutoring sessions to improve students' engagement (e.g., let students work on problems or let them explain a concept to you)? How and when?
    \item Did you give prompt and positive feedback to students to improve students' engagement?(e.g., when they make progress towards solving a problem or when they show good behavior)? How and when?
    \item Did you encourage students to be persistent to improve students' engagement(e.g., don't allow students to give up or scaffold until they succeed)? How and when?
    \item Did you give autonomy to your students to improve students' engagement (e.g., let students participate when making decisions)? How and when?
    \item Did you set time constraints for activities inside your tutoring sessions to improve students' engagement (e.g., asking students to solve a practice problem in 5 minutes)? How and when?
    \item Did you present yourself to your students as a positive role model to improve students' engagement (show a growth mindset or start sessions on time)? How and when?
\end{enumerate}

\end{document}